\documentclass[11pt ]{article}
\usepackage[affil-it]{authblk}

\usepackage{graphicx}
\usepackage{amsmath, amsthm, amssymb}
\usepackage{subfigure}
\usepackage{comment}
\usepackage{bm}
\usepackage{epsfig,color}
\usepackage[sans]{dsfont}
\usepackage{verbatim}
\usepackage{amsfonts}
\usepackage{amsmath}
\usepackage{amssymb}
\usepackage{enumerate}

\newcommand{\tr}{\text{tr}}

\newcommand{\be}{\begin{equation}}
\newcommand{\ee}{\end{equation}}
\newcommand{\bea}{\begin{eqnarray}}
\newcommand{\eea}{\end{eqnarray}}
\newcommand{\bes}{\begin{equation*}}
\newcommand{\ees}{\end{equation*}}
\newcommand{\beas}{\begin{eqnarray*}}
\newcommand{\eeas}{\end{eqnarray*}}

\newcommand{\x}{\mathrm{x}}

\renewcommand{\H}{\mathcal{H}}

\def\B{\mathcal{B}}

\def\x{\mathrm{x}}

\def\g{\mathrm{guess}}

\def\tr{\mathrm{tr}}

\newtheorem*{thm*}{Theorem}

\newtheorem*{lem*}{Lemma}

\newtheorem*{lipschitzLem*}{Lemma \ref{lipschitz}}
\newtheorem*{lipschitzCubeLem*}{Lemma \ref{lipschitzCube}}
\newtheorem*{pgmNearlyOptimalThm*}{Theorem \ref{pgmNearlyOptimal}}
\begin{document}

\title{Quantum State Discrimination and Its Applications}

\author{Joonwoo Bae$^{1,2}\footnote{bae.joonwoo@gmail.com}$ and Leong-Chuan Kwek$^{2,3 ,4}\footnote{kwekleongchuan@nus.edu.sg}$}

\affil{
$^{1}$ Freiburg Institute for Advanced Studies (FRIAS), Albert-Ludwigs University of Freiburg, Albertstrasse 19, 79104 Freiburg, Germany \\
$^{2}$ Center for Quantum Technologies, National University of Singapore, \\ 3 Science Drive 2,  Singapore 117543 \\
$^{3}$ National Institute of Education, 1 Nanyang Walk, Singapore   637616 \\
$^{4}$ Institute of Advanced Studies, Nanyang Technological University, \\ 60 Nanyang View, Singapore 639673 }

\maketitle

\begin{abstract}

Quantum state discrimination underlies various applications in quantum information processing tasks. It essentially describes the distinguishability of quantum systems in different states, and the general process of extracting classical information from quantum systems. It is also useful in quantum information applications, such as the characterisation of mutual information in cryptographic protocols, or as a technique to derive fundamental theorems in quantum foundations.  It has deep connections to physical principles such as relativistic causality.  Quantum state discrimination traces a long history of several decades, starting with the early attempts to formalise information processing of physical systems such as optical communication with photons. Nevertheless, in most cases, optimal strategies of quantum state discrimination remain unsolved, and related applications are valid in some limited cases only. The present review aims to provide an overview on quantum state discrimination, covering some recent progress, and addressing applications in some selected topics. This review serves to strengthen the link between results in quantum state discrimination and quantum information applications, by showing the ways in which the fundamental results are exploited in applications and vice versa. 

\end{abstract}

\section{Introduction}
 
Quantum information science arises from the need to seek for faster and more efficient means of processing information and computation. However, in order to exploit quantum mechanical features of photons, atoms and molecules for information processing, a deeper understanding of the physics behind these microscopic systems is essential. 

The information embedded in quantum mechanical systems are fully described by their quantum states.
Unlike their classical counterparts, unknown quantum states cannot be copied perfectly\cite{ref:bb84, ref:no-cloning} (no-cloning theorem). Moreover, by exploiting the quantum entanglement of these quantum states, it is possible to obtain quantum correlations with no classical equivalence \cite{ref:bbm}. Entangled quantum states violate  local realistic descriptions of Nature and cannot be described by any hidden variable theories \cite{ref:e91, ref:bell}.  However, these intriguing features and properties of quantum mechanical states can be harnessed for various cryptographic protocols giving rise to the whole field of quantum cryptography. Indeed, the no-cloning theorem discovered in 80's lends further weight to the argument that exploiting quantum communication would enhance the level of security \cite{ref:bb84}.

In fact, the fundamental properties that make such quantum applications possible are also closely related to the fact that non-orthogonal quantum states cannot be discriminated perfectly. If quantum state discrimination were perfect, it would lead directly to the contradiction that quantum cloning could be done perfectly, or that quantum entanglement could lead to instantaneous communication \cite{ref:gisin98}.  Thus, state discrimination lies at the heart of some of these quantum applications . 

The indistinguishability of quantum states is also implicit within the postulates of quantum theory. Indeed, long before the interest in quantum information started, it was found that optimal quantum state discrimination as well as measurement strategies could be studied rigorously and cast into firm mathematical foundation\cite{ref:holevo, ref:yuen, ref:hel}. Historically, these studies were motivated by the development of optical communication in the seventies. 

The best strategy adopted for quantum state discrimination depends largely on the figure of merit used, for instance, see reviews \cite{ref:rev1, ref:rev2, ref:rev3, ref:rev4, ref:rev5}. In some applications, one may wish to minimize the average errors that occur in a state discrimination task, and in others, one would like to maximised the confidence level of detection events in the measurement. Different strategies adopted may lead to different outcomes. Nonetheless,  there are always consistencies amongst these strategies, e.g. the the strategy for maximum confidence level in state discrimination is really equivalent to the strategy for the minimum-error discrimination if an average over the given states is taken, or to the strategy for unambiguous state discrimination if the given states are linearly independent. 

In recent years, the need for quantum state discrimination appears widely in quantum information processing  and also in the study on the foundation of quantum theory. For instance, it is a key tool to obtain a no-go theorem for an interpretation of quantum states \cite{ref:pbr}. In addition, the characterisation of the indistinguishability of quantum states has also been studied from a more fundamental perspective, in the so-called $\psi$-epistemic theories \cite{ref:leifer} and generalised probabilistic theories \cite{ref:bae11}.  Quantum state discrimination is also needed for the search for a dimension witness of quantum systems \cite{ref:dimqsd} as well as an operational interpretation of conditional mutual entropy \cite{ref:ckr}.  

Optimal quantum state discrimination is generally difficult apart from the case of two-state discrimination. However,  there has been much progress in recent years towards the subject.  For instance, an analytic solution for the discrimination of qubit states assigned at random is available. 

The goal of this current review is  to provide a fairly comprehensive introduction to quantum state discrimination and its various applications. It is hoped that some of these new results of state discrimination may lead to new insights in other issues in quantum information science and vice versa. In Sec. \ref{sec:overview}, we first review various methods of quantum state discrimination and the recent progress along the line. In particular, there are a number of applications based on two-state minimum-error discrimination. In Sec. \ref{sec:tool}, we cover selected results obtained recently in quantum information theory where state discrimination has been exploited as a major tool, such as in interpretation of quantum states in foundation and as dimension witness for device-independent quantum information processing. In Sec. \ref{sec:characterization}, we address contexts where optimal state discrimination is characterized by information-theoretic quantities and also physical principles. Finally, in Sec. \ref{sec:conclusion}, we summarize our report and make a few concluding remarks.


\section{Discrimination of Quantum States}
\label{sec:overview}

Quantum state discrimination can be regarded as a game involving two parties, typically called Alice and Bob: Alice first prepares a quantum state and send it to Bob and then Bob applies appropriate measurement. An important assumption in this exchange is that the two parties make prior arrangement concerning the set of quantum states.   Bob's goal is to determine which state has been prepared by Alice knowing that the state receieved is taken from an {\it a priori} agreed set of quantum states.  The question now  is how much one can learn about the state prepared through quantum measurement together with the classical knowledge available from  {\it a priori} information.   This is a non-trivial problem since any measurement on a quantum system does not always reveal full information about the prepared system. Arbitrary set of quantum states need not be orthogonal and so they are generally not distinguishability from each other. 

It is worth emphasising that indistinguishability of quantum states is not due to any  lack of knowledge as in classical systems. Classical states are perfectly distinguishable from one another since they can be identified unambiguously. Indistinguishable classical states only mean that both preparation and measurement are described,  due to the lack of knowledge, by probability distributions that overlap one another. Recently, it is shown that indistinguishability of quantum states cannot be explained by overlapping probability distributions of classical states \cite{ref:leifer}. In short, indistinguishability of quantum states does not have a classical analogue. 

Since quantum states cannot be discriminated perfectly, one naturally seeks an alternative option: by adopting a figure of merit, the discrimination scheme is designed and the measurement setting is then optimised. There are a number of figures of merit in quantum state discrimination, depending the application needed. For instance, measurements are applied in single-shot manner in which the states are measured one at a time as soon as they arrive, or performed repeatedly or adaptively on independent and identically distributed state preparation. Alternatively, the average error incurred in the discrimination scheme is minimised. Or, incorporating inconclusive outcomes, the measuring party can unambiguously determine the prepared state. In this section, we review some of these methods for discriminating quantum states and their recent progress. 

\subsection{ A basic setting } 
\label{subsec:setting}

\begin{figure}[t]
\begin{center}
\includegraphics[width=4.7in]{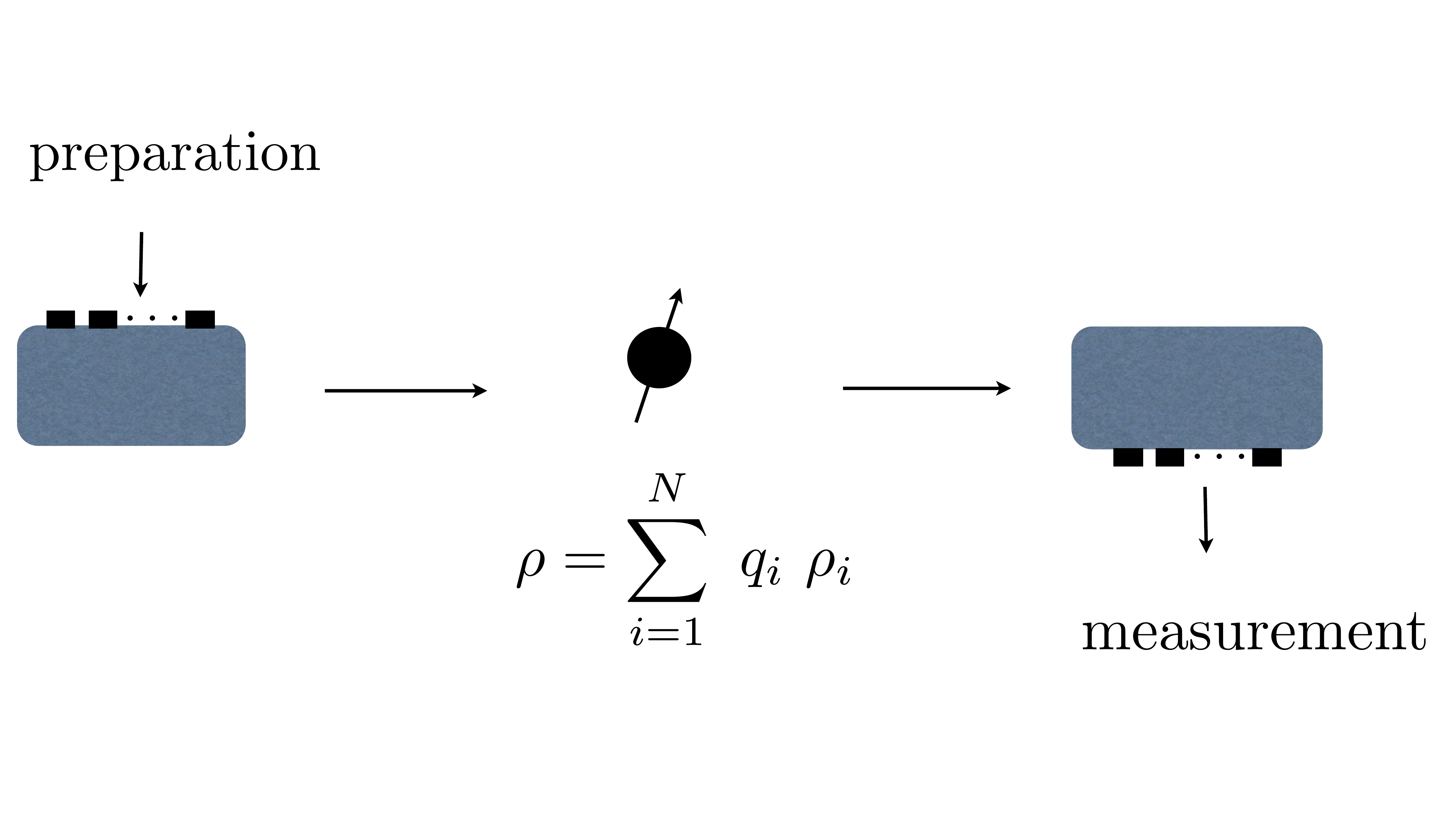}
\caption{Two devices for preparation and measurement, respectively, are the building blocks in implementation of quantum state discrimination. The device for preparation generates a state $\rho_i$ with { \it a~priori } probability $q_i$ for some $i\in [1,N]$, and sends it to the other party for measurement. Once a system is generated and before measurement happens, the system is described by the probabilistic mixture of all possibilities. It can be described as $\rho = \sum_{i=1}^N q_i \rho_i $. Then, measurement is applied and produces a detection event, from which one can learn about the preparation. }
\label{fig1:scenario}
\end{center}
\end{figure}

Let the  state be taken from a $d$-level quantum system be denoted by a unit-trace positive operator $\rho$ on a $d$-dimensional Hilbert space $\H_d$. Let us also write the set of quantum states by $\mathcal{S}(\mathcal{H}_d)$. In short, the states, $\rho\in \mathcal{S}(\mathcal{H}_d)$, satisfies $\rho > 0$ and $\mathrm{tr} \rho = 1$ on Hilbert space $\H_d$. For the preparation, let $\{ \rho_i\}_{i=1}^N$ denotes the set of {\it a priori} quantum states that Alice and Bob have agreed at the beginning and $\{q_i \}_{i=1}^N$ be the corresponding {\it a priori} probabilities, i.e. $\{ q_i , \rho_i \}_{i=1}^{N}$ as the preparation process in which one of states $\{ \rho_i\}_{i=1}^{N}$ is generated with {\it a priori} probability $\{q_i\}_{i=1}^N$ respectively.  Since one of the quantum states from the set  is generated during the preparation step, it follows that $\sum_{i=1}^N q_i = 1$.  If we do not know which quantum state is being produced, the state after the preparation process is described as a mixed state 
\bea
\{q_i , \rho_i  \}_{i=1}^N:~~~~\rho = \sum_{i=1}^N q_i \rho_i,~~\mathrm{with}~~\sum_{i=1}^N q_i =1. \label{eq:state1}
\eea
We depict the preparation scenario in Fig.(\ref{fig1:scenario}), where Alice presses a button with the {\it a priori} probability for generating her choice of a state from $\{\rho_i \}_{i=1}^N$.

Bob's device for detecting Alice's state should generally have  $L$ output ports with $L \geq N$ i.e. no less than the number of possible states in the preparation. Depending on  the figures of merit used, there are different optimal strategies for the measurement process. Quantum measurements are described by positive-operator-valued-measure (POVMs), or also called as probability-operator measurement (POM), which are positive operators $M_{k} \geq 0$ for $k=1,\cdots,L$ that fulfill $\sum_{k=1}^L M_k = I$, i.e. a positive resolution of the identity operator.  We should note that one can always find a method of realising such POVMs - at the very least, one may invoke Naimark's extension, which says the extra resources can be involved as ancillary systems and projective measurements on the extra resources can provide a means to construct POVMs on given systems. In fact, it is interesting to study the behaviour of correlations between systems and ancillary systems, see also Refs. \cite{roa2011dissonance} and \cite{zhang2013requirement}.

One can regard each POVM  as an output port where a particular detection outcome happens. For instance, let $M_k$ denote the POVM corresponding to the $k$th port, then the detection event occurs with some probability, say $p(k)$ depending on how the measurement setting $\{ M_k\}_{k=1}^N$ is constructed. In general, for a state $\rho_i$ generated in the preparation, detection event on $M_k$ happens with probability given by
\bea 
p(k | i ) = \mathrm{tr} [ {M_k \rho_i} ],~~\mathrm{for}~k=1,\cdots,L ,~\mathrm{and}~ i=1,\cdots, N.\label{eq:msprob}
\eea
One can also check the sum of probabilities is still unity, i.e. $\sum_{k=1}^{L} p(k | i) =1$ since $\sum_{k=1}^{L} M_k = I$. 

We next review some known methods for distinguishing quantum states, namely the minimum-error discrimination, unambiguous state discrimination, and maximum confidence discrimination. Note that we here focus and spend more on minimum-error state discrimination, that has been much developed recently, and refer to other excellent reviews Refs. \cite{ref:rev1, ref:rev2, ref:rev3, ref:rev4, ref:rev5} for the others such as details of unambiguous discrimination and maximum confidence measurement, as well as some of experimental progress.

\subsection{ Minimum-error discrimination}
\label{subsec:med}

The goal here is to minimise the average error (or  maximise the success probability for the guess). The maximum success probability is called the guessing probability. During measurement, if a detection event happens at the $k$th output port, Bob concludes that Alice has prepared state $\rho_k$.  Such an assignment cannot be perfect unless states $\{ \rho_i\}_{i=1}^N$ are pairwise orthogonal. Otherwise, an error is incurred in the process.

\begin{figure}[t]
\begin{center}
\includegraphics[width=4.7in]{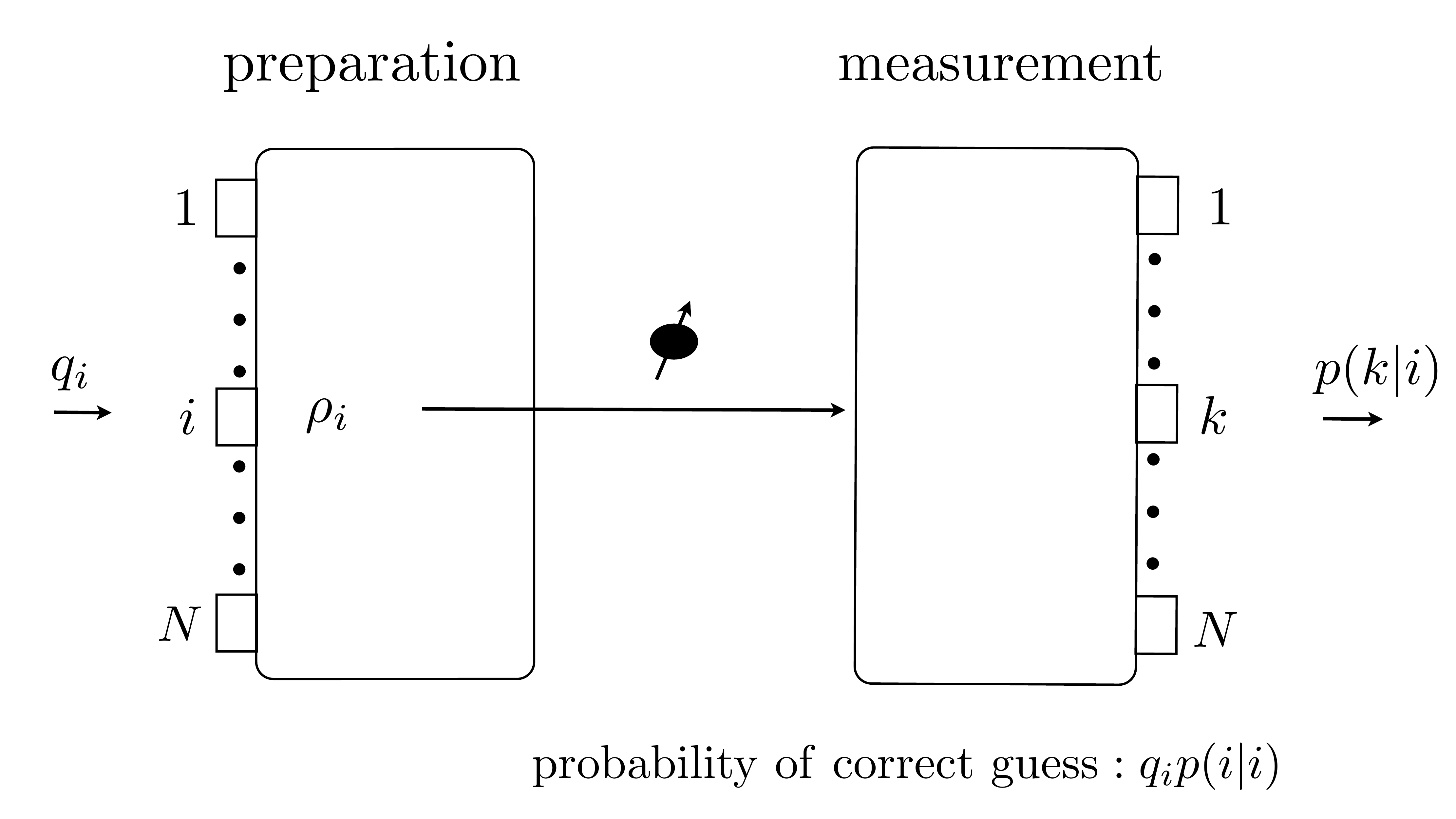}
\caption{ In minimum-error discrimination, there are $N$ inputs and $N$ outputs. During the preparation stage, one presses a button $i$ among $1,\cdots, N$ with probability $q_i$. This generates state $\rho_i$, which is sent to the measurement device. In the measurement device, there are $N$ output ports, one of which must show a detection event since a complete measurement is applied. Given that $\rho_i$ is generated, an output port $k$ shows a detection event with probability $p(k|i)$. Overall, the probability that one makes a correct guess once an output port $i$ is clicked is given by $q_i p(i | i)$.}
\label{fig2:meqsd}
\end{center}
\end{figure}

As quantum measurement is described by POVMs, this optimisation over measurement devices requires a search for optimal and experimentally viable POVMs that can give rise the minimum error.  We therefore seek for POVMs $\{M_k \}_{k=1}^N$ so that the detection event on each $M_k$ leads to the state $\rho_k$  with minimum average error:
\bea
\mathrm{detection ~event ~ on ~ }M_k ~ \rightarrow ~ \rho_k   ~ \mathrm{is~ given}. \label{eq:mind}
\eea
When the state $\rho_i$ is sent to the measurement device, the probability that a detection event occurs  corresponding to the POVM $M_i$ is given by $\tr[M_i \rho_i]$. Since each state $\rho_i$ appears with $a ~priori$ probability, the probability of guessing a state $\rho_i$ correctly is given by $q_i p( i  | i ) = \mathrm{tr} [ M_i q_i \rho_i ]$, see Figure (\ref{fig2:meqsd}). The guessing probability is then obtained by maximising the success probability of correct guesses: 
\bea
p_{\g} & = & \max_{\mathrm{M }} \sum_{i=1}^N q_i p(i|i) =  \max  ~\sum_{i=1}^N  q_i \tr[M_i \rho_i ]  \nonumber \\
 && \mathrm{subject~to }~~\sum_i M_i = I , ~~ \forall~M_i\geq 0, \label{eq:medprob}
\eea
where $\mathrm{M}$ denotes measurement. The minimum average probability that one fails to guess correctly a state is then $p_{\mathrm{error}} = 1-p_{\g}$. Note that Eq. (\ref{eq:medprob}) does not provide a simple recipe for arriving at the guessing probability. Moreover, in general, even if the guessing probability is obtained, it is still non-trivial to find the optimal measurements or its general properties. In short, the main task in minimum-error discrimination is to find both the optimal guessing probability and  measurement. 

The optimal measurement for minimum-error discrimination for a set of quantum  states $\{ q_i ,\rho_i \}_{i=1}^N$  is in general not unique.  Different sets of POVMs can give rise to the same guessing probability \cite{ref:mochon}. In some cases, a trivial guess can be made  with just  {\it a priori} knowledge and without further measurement \cite{ref:hunter}.  In other cases,  an optimal strategy may  contain zero-valued POVM i.e. $M_i =0$ for some states $\rho_i$ so that there is no output port associated with the measurement setting.   From the point of experimental implementation, such a strategy reduces the experimental resources and specifies the minimum number of  output ports associated with non-zero POVM elements. Note that the non-zero POVM in optimal measurement must be projectors \cite{ref:hel, ref:yuen}.  An optimal strategy for minimum-error discrimination is in general not easy to find, but in some specific applications, such an optimal discrimination strategy is readily obtained. As a general method,  numerical optimisation are exploited, for instance, semidefinite programming \cite{ref:sdpbook}. This is feasible as long as the numerical method yields solutions within  polynomial time.  

\subsubsection{Two-state discrimination}

Discrimination between two quantum states is probably the only case where optimal discrimination is known without any imposing further assumptions\cite{ref:hel}. Suppose that the two quantum states are chosen from the set  $\{q_i ,\rho_i\}_{i=1}^2$, and there are two output ports denoted by $\{ M_1 , M_2 \} $. A detection event on each port $M_i$ for $i=1,2$ corresponds to a state $\rho_i$. Each state is prepared according to an {\it a priori} probability $q_i$ for $i=1,2$ and the probability of making a correct guess  is given by $q_i p(i|i)$, see also Eq. (\ref{eq:mind}). The guessing probability is obtained by maximising the average success probability over all possible measurements, 
\bea
p_{\g} = \max_{\mathrm{Measurement}}  q_1 p(1 | 1) +  q_2 p(2 |2 )  =  \tr [ K  ], ~\mathrm{with}~K =  q_1  M_1 \rho_1 + q_2  M_2 \rho_2, \nonumber 
\eea
where we have introduced an operator $K$, determined by optimising the POVMs. Since  $M_1 + M_2 = I$, the operator $K$ can be shown to be
\bea
K  = \left\{ \begin{array}{l}
 q_2 \rho_2 + M_1 X  \\ 
 q_1 \rho_1 - M_2 X,~~~  \mathrm{where} ~~X = q_1\rho_1 - q_2\rho_2.
\end{array} \right.
\eea
By symmetrising the above expression, we get
\bea
K = \frac{1}{2} (q_1 \rho_1 + q_2 \rho_2) + \frac{1}{2}(M_1 - M_2 ) X.  \nonumber 
\eea
If we denote  $M = M_1  - M_2$, we see that the POVMs $M_1 = (I +M )/2$ and $M_2 = (I - M)/2$, both of which must be non-negative, and the guessing probability can be obtained through the optimisation of a single parameter $M$ satisfying $-I \leq M \leq I$,
\bea
p_{\g} = \max_{M} \tr[K] =\frac{1}{2} + \max_M \frac{1}{2} \tr M X. \label{eq:guess2}
\eea
This optimal $M$ can be found by spectral decomposition of operator $X = q_1\rho_1 - q_2\rho_2$ into positive and negative parts. 

Let us write the spectral decomposition of the operator by $X = \lambda_{+} X_{+} - \lambda_{-} X_{-}$ with positive (negative) projector $X_{+}$ ($X_{-}$) and $\lambda_{+}$ ($\lambda_{-}$) positive (negative) eigenvalue. This is always possible to do with given states $\{q_i,\rho_i \}_{i=1}^2$. The optimal choice is given by $M = X_{+} - X_{-}$ with the guessing probability $p_{\g} = (1+ \lambda_{+} + \lambda_{-})/2$. In fact, in terms of the $1$-norm, 
\bea
\| X  \|_{1} = \sum_{i=\pm} | \lambda_{i} | = \tr {| X |} = \tr \sqrt{ X^{\dagger} X}. \label{eq:1norm}
\eea
The guessing probability is then rewritten as,
\bea
p_{\g} = \frac{1}{2} +  \frac{1}{2} \|  q_1 \rho_1 - q_2 \rho_2 \|_1.  \label{eq:helbd}
\eea 
This is also known as the Helstrom bound \cite{ref:hel}.  From the optimal choice of $M$, the optimal POVMs are
\bea
M_1 = \frac{1}{2} (I + M) = X_{+},~\mathrm{and} ~M_{2} = \frac{1}{2} (I - M) = X_{-}, \label{eq:optm2}
\eea
which form a projective measurement. It is also worth noting that the derivation above is strictly valid only for any two given quantum states. 

\subsubsection{ Multiple state discrimination  }
\label{subsec:multi}

For more than two states, optimal state discrimination is only solved in some limited cases where the given states possess certain symmetries or the dimension of the Hilbert space of the quantum states is small. For instance, optimal discrimination of multiple states is known for arbitrary qubit states that are given equal {\it a priori} probabilities, or for "geometrically uniform" states. In this subsection,  we review optimal discrimination of geometrically uniform states and mirror-symmetric states. 

\begin{figure}[t]
\begin{center}
\includegraphics[width=5.2in]{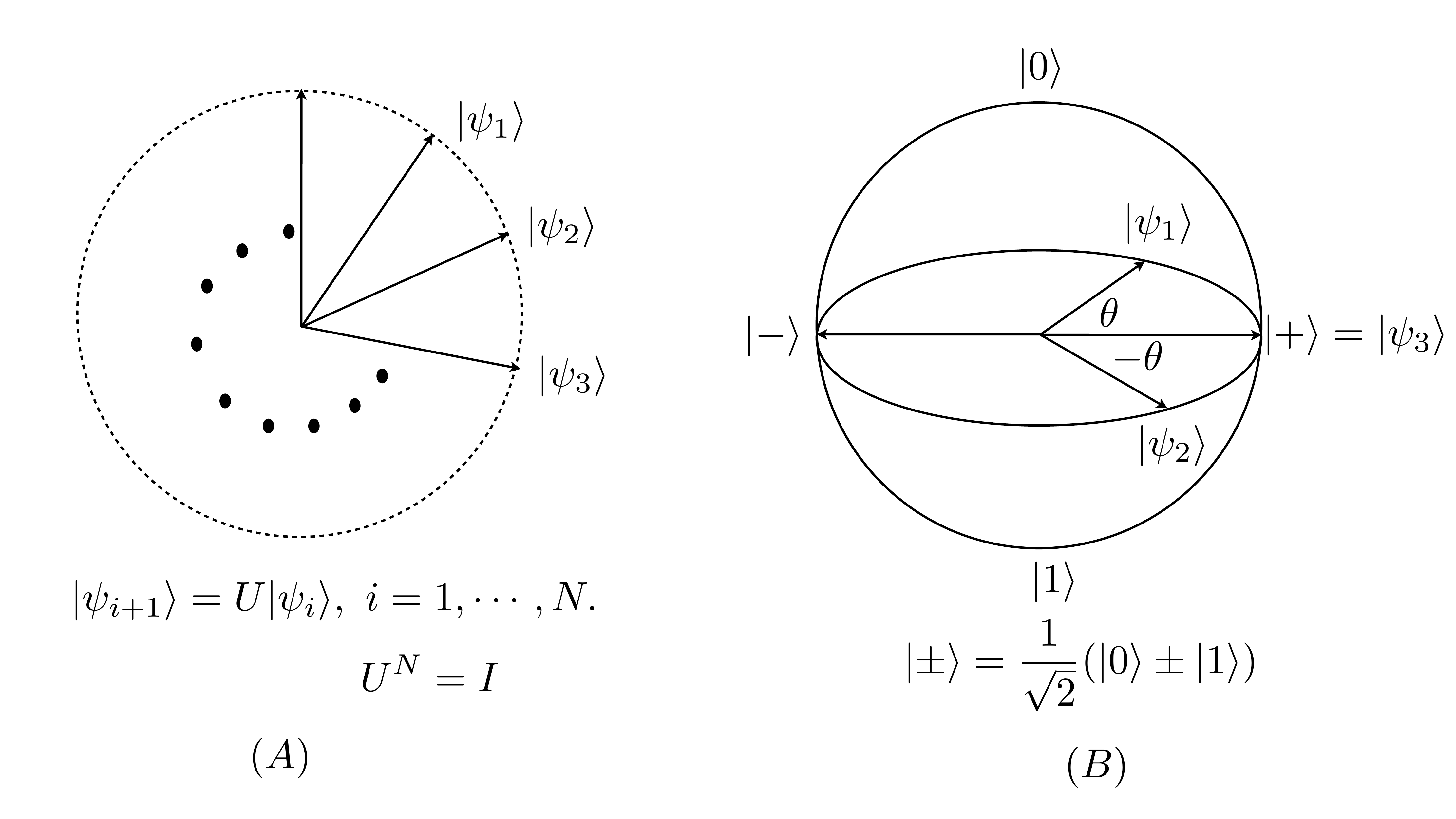}
\caption{ (A) Geometrically uniform states can be understood as a set of states forming a circle connected by a single unitary transformation $U$. There is a cyclic symmetry such that $U | \psi_i \rangle = |\psi_{i+1} \rangle$ with $U^{N} = I$.  (B) Mirror-symmetric states in Eq. (\ref{eq:misyms}) are shown in the Bloch sphere. Note that two states $|\psi_{1} \rangle$ and $|\psi_2 \rangle$ are obtained from state $|\psi_3 \rangle = | + \rangle$, by rotating it with angle $\theta$ in opposite directions, respectively.}
\label{fig3:fig3}
\end{center}
\end{figure}

\noindent{\it Geometrically uniform states}. Quantum states $\{\rho_i  \}_{i=1}^N$ where $\rho_i \in S(\H_d)$ are called geometrically uniform \cite{ref:eldar-sdp}, see Fig. (\ref{fig3:fig3}) if there exists a symmetry represented by  unitary transformations $U$ such that each state $\rho_i$ is transformed to $\rho_{i+1}$ by an application of the unitary for all $i=1,\cdots, M$: 
\bea
\rho_{i+1} = U \rho_i U^{\dagger} ~~\mathrm{for~all }~ i = 1,\cdots,M ~~\mathrm{with ~} U^{M} = I~\mathrm{and}~\rho_{M+1} = \rho_1. \label{eq:geostate}
\eea
To see how this symmetry can be exploited for state discrimination, let us consider $d$ geometrically uniform states in a $d$-dimensional Hilbert space $\H_d$ assuming the equal {\it a priori} probability $q_\alpha = 1/d$ for all $\alpha=0,\cdots, d-1$. In particular, we consider the following $N$ copies of pure states 
\bea
 \rho_\alpha &=& |\psi_\alpha \rangle \langle \psi_\alpha|^{\otimes N},~\mathrm{for}~\alpha=0,\cdots,d-1, ~\mathrm{where}~ | \psi_\alpha \rangle = \sum_{n = 0 }^{d-1}  c_n \exp ( \frac{2\pi i }{d} n \alpha) | x_n \rangle,~
 \label{eq:geoex}
\eea 
where  $\{ |x_n \rangle \}_{n=0}^{d-1}$ is an orthonormal basis for Hilbert space $\H_d$. These states are of particular interest as they have been considered in an optimal eavesdropping strategy in quantum key distribution \cite{ref:jbae2way}. The unitary transformations that finds the symmetry among the states is given by $U^{\otimes N}$ where 
\bea
U = \sum_{m=0}^{d-1} \exp(\frac{2\pi i }{d}m ) |x_m \rangle \langle x_m |,~\mathrm{such~that}~ U| \psi_\alpha\rangle = |\psi_{\alpha+1} \rangle,~ \forall\alpha \label{eq:unisym}
\eea
With this unitary transform, one can find that $(U^{\otimes N} )\rho_{\alpha} (U^{\otimes N})^{ \dagger} = \rho_{\alpha+1}$ for all $\alpha=0,\cdots,d-1$.

Given the symmetry, the eigenvalues of an equal mixture of the states, $\rho= \sum_\alpha \rho_{\alpha}/d$, can be computed in an analytic way. For convenience, let us consider the case of $N=1$ for now, and the properties apply to the other cases of $N>1$. For the geometrically uniform states, one can find orthogonal basis spanning the support of those states explicitly as
\bea
|x_n\rangle = \frac{1}{d c_n} \sum_{\alpha} \exp(- \frac{2\pi i }{d} n \alpha) |\psi_{\alpha} \rangle. \nonumber
\eea
Then, with equal probabilities $q_{\alpha} = 1/d$, it holds that
\bea
\rho = \frac{1}{d} \sum_{\alpha} |\psi_\alpha \rangle \langle \psi_{\alpha} | = \sum_{n} c_{n}^2 |x_n \rangle \langle x_n | \nonumber
\eea
This simply shows a spectral decomposition of the state $\rho$ with eigenvalues $ \{ \lambda_{\mu} \}_{\mu = 0 }^{d=1}$
\bea
\lambda_\mu =  c_{\mu}^2  = \frac{1}{d^2} \sum_{\alpha,\beta} \exp(\frac{2\pi i }{d}  \mu (\beta - \alpha) ) \langle \psi_\beta | \psi_\alpha \rangle, \nonumber
\eea
which are useful to find optimal discrimination. Note also a useful relation that $\langle \psi_\beta | \psi_\alpha \rangle = \langle \psi_{(\alpha-\beta) } |\psi_0 \rangle$. Then, following the method in Ref. \cite{ref:eldar-sdp}, one then constructs a matrix $\Phi = \sum_{\gamma} |\psi_{\gamma} \rangle \langle x_{\gamma}|$ and obtains the spectral decomposition for the optimal measurement. In fact, for the above geometrically uniform states, the matrix $\Phi$ is diagonalised by a Fourier transformation, $\mathcal{F} | x_n \rangle = \sum_{w} \exp(\frac{-2\pi i}{d} wu) / \sqrt{d}|x_w  \rangle$. The optimal POVM are $\{ M_{j} = | m_j \rangle \langle m_j |\}_{j=0}^{d-1} $ where
\bea
|m_j\rangle = \frac{1}{\sqrt{d}} \sum_{k=0}^{d-1} \exp(\frac{2\pi i }{d} jk) |x_k \rangle. \nonumber 
\eea
Note that the measurement is really an example of the square-root-measurement (or also called as ``pretty good" measurement) \cite{ref:ban, ref:sasaki, ref:hau, ref:mochon}. In this case, optimal POVMs fulfill the relation that $M_j \propto \rho^{-1/2 } | \psi_j \rangle \langle \psi_j |  \rho^{-1/2}$ with the state preparation $\rho = \sum_{i} q_i \rho_i$.  With the optimal measurement, it is straightforward to compute the guessing probability,
\bea
p_{\g} = \sum_{j=0}^{d-1}  q_j |  \langle  m_j |  \psi_j  \rangle  |^2  =   \frac{1}{d} |  \sum_{\mu} c_{\mu}  |^2 = 
\frac{1}{d^2} \big|  \sum_{\eta} (  \sum_{m} \exp( \frac{ 2\pi i}{d} \eta m)  \langle \psi_m  | \psi_0 \rangle  )^{1/2}   \big|^2. \nonumber
\eea

Coming back to the original consideration for states $\{ |\psi_{\alpha}\rangle^{\otimes N} \}_{\alpha=0}^{d-1}$ with $N>1$,  the properties of geometrical uniform states in the above are also applied. These are geometrically uniform states with the symmetry given by the unitary transformation $U^{\otimes N}$ with $U$ in Eq. (\ref{eq:unisym}). Then, it is straightforward to compute the guessing probability as, 
\bea
p_{\g} = \frac{1}{d^2} \big|  \sum_{\eta} (  \sum_{m} \exp( \frac{ 2\pi i}{d} \eta m)  \langle \psi_m  | \psi_0 \rangle^{N}  )^{1/2}   \big|^2. \nonumber
\eea
It is worth emphasizing that it is the inherent symmetry facilitates the computation of the guessing probability and the optimal measurement. \\

\noindent{\it Mirror-symmetric states}. Optimal discrimination has also been obtained for  mirror-symmetric states. These are linearly dependent three states lying on an equator of the Bloch sphere
\bea
|\psi_{1} \rangle &=& \cos\theta |+\rangle + \sin\theta | - \rangle \nonumber\\
|\psi_{2} \rangle &=& \cos\theta |+\rangle - \sin\theta | - \rangle \nonumber\\
|\psi_{3} \rangle &=& | + \rangle \label{eq:misyms}
\eea
where two orthogonal basis are denoted by $|\pm\rangle = ( |0\rangle \pm |1\rangle)/ \sqrt{2}$.  These states are considered in Ref. \cite{ref:mirror1} with  {\it a priori} probabilities given by $q_1 = q_2  = p$ and $q_3 = 1-2p$.  Mirror symmetry refers to the symmetry forms when the state $|\psi_3\rangle$ is roated in opposite direction through the same angle $\theta$.
The state preparation $\rho = \sum_{i} q_i \rho_i$ is invariant under the transformation $|+ \rangle \rightarrow |+ \rangle$ and $| -\rangle \rightarrow - |-\rangle$. The guessing probability is obtained as
\bea
\mathrm{if}~~p \geq p_{*},  ~~  p_{\g} & = & p (1 + \sin2\theta),\nonumber \\
\mathrm{if}~~p \leq p_{*},  ~~  p_{\g} & = & \frac{(1-2p)(p\sin^2\theta + 1 -2p -p\cos^2\theta)  }{ 1-2p -p\cos^2 \theta} \label{eq:gmirror}
\eea
where $p_* = (2 + \cos\theta (\cos\theta + \sin\theta)  )^{-1}$. Note that the optimal measurement provided in Ref. \cite{ref:mirror1} is not a square-root-measurement, nor does it attain the maximum guessing probability. Therefore, the example also shows that square-root-measurement is not a general optimal strategy for minimum-error discrimination. 

\subsubsection{ A general approach}
\label{sec:generalapp}

Although minimum-error discrimination was formalised several decades ago \cite{ref:hel, ref:holevo, ref:yuen}, analytical solution of optimal discrimination is only known in some specific cases.  One such case is optimality condition which characterise optimal measurement and the guessing probability in general. Note that they were already obtained in earlier works by Holevo and Yuen \cite{ref:yuen, ref:holevo, ref:hel}. 

Recently, a general approach for optimal state discrimination has been addressed by exploiting quantum state geometry\cite{ref:baenjp}. The approach has been derived from the insight gained from the convex optimisation as a linear complementarity problem, more details can be found in Ref. \cite{ref:baenjp}.

{\it Optimality conditions}. The optimality conditions are a set of necessary and sufficient conditions for POVMs to be optimal measurements. For the given states $\{q_i ,\rho_i \}_{i=1}^N$, the POVM elements $\{M_i \}_{i=1}^N$ provide a set of optimal measurements if and only if they satisfy  \cite{ref:yuen,ref:holevo, ref:hel}
\bea
M_i (q_i \rho_i - q_j \rho_j) M_j & =& 0 ~\forall~i,j=1,\cdots,N, \label{eq:oc1} \\
\sum_{i=1}^{N} q_i \rho_i M_i - q_j \rho_j & \geq &0 ~\forall~j=1,\cdots,N, \label{eq:oc2}   
\eea
These conditions can also be applied to a numerical optimisation scheme \cite{ref:jez} where it is shown that the optimality conditions can be rewritten as, 
\bea
q_i \rho_i + r_i \sigma_i  &=& K,~ \forall~ i=1,\cdots, N, \label{eq:medopt3}\\
r_i \tr [\sigma_i M_i] & = & 0 ~\forall~ i=1,\cdots, N. \label{eq:medopt4}
\eea 
with the complementary states $\{\sigma_i\}_{i=1}^N$, coefficients $\{ r_i \geq 0 \}_{i=1}^N$, and a positive Hermitian operator $K$ \cite{ref:baenjp}. The two forms of optimality conditions are equivalent \cite{ref:baenjp}.  The optimal POVMs in Eq. (\ref{eq:medopt4}) are orthogonal to corresponding complementary states. The guessing probability can be found from Eq. (\ref{eq:medopt3}) so that, with the optimal POVMs $\{ M_i \}_{i=1}^N$, it follows that
\bea
p_{\g} = \tr [ \sum_i q_i  \rho_i M_i ] = \tr [ \sum_i  ( q_i \rho_i  + r_i \sigma_i) M_i   ] = \tr [ K \sum_i M_i ] = \tr[K ]. \label{eq:geog}
\eea
In particular, $K$ is the symmetry operator in that it is uniquely determined for a set of states even though the optimal POVMs are in general not unique. Moreover, the optimal discimination is achieved once $K$ is obtained for given states.

The optimality conditions in Eqs. (\ref{eq:medopt3}) and (\ref{eq:medopt4}) are not obtained directly from Eqs. (\ref{eq:oc1}) and (\ref{eq:oc2}). The conditions are derived from a linear complementarity problem in the context of convex optimisation.  In convex optimisation, optimal state discrimination is described by the following optimisation problem: 
\bea
(\mathrm{Primal})~~&&  \max  \sum_{i=1}^N q_i \tr[M_i \rho_i ] \nonumber \\
&& \mathrm{subjec ~ to}~  \sum_{i} M_i = I,~~~  M_i  \geq 0,~~ i =1,\cdots, N \label{eq:primal}
\eea
or, equivalently
\bea
(\mathrm{Dual})~~&&  \min   \tr[K] \nonumber \\
&& \mathrm{subjec ~ to}~  K \geq q_i \rho_i~~i = 1,\cdots,N. \label{eq:dual}
\eea
The former one is called a primal problem in the optimisation and the latter  is its  dual. In general, the two problems may not yield the same solution. When the condition of strong duality is satisfied, both problems give the same solution, i.e. the guessing probability can be obtained by solving any one of them. 

There is an approach called linear complementarity problem (LCP) that generalises primal and dual problems:
\bea
(\mathrm{LCP})~~&&  K = q_ i \rho_i + r_i \sigma_i,~~\mathrm{for}~i=1,\cdots, N~~ (\mathrm{Lagrangian ~ stability}), \nonumber \\
&&  r_i \tr[\sigma_i M_i] = 0,~~ \mathrm{for}~i=1,\cdots, N ~~ (\mathrm{complementarity ~slackness}),  \nonumber \\
&& \sum_{i} M_i = I,~~~  M_i  \geq 0,~~ i =1,\cdots, N \nonumber \\
&&  K \geq q_i \rho_i~~i = 1,\cdots,N. \label{eq:lcp}
\eea
These conditions are also known as the Karush-Kuhn-Tucker conditions. In this approach, the task is not really an optimisation problem but rather it is a search for suitable parameters, operator $K$ and $\{ \sigma_i \}_{i=1}^N$, and constants $\{ r_i \}_{i=1}^N$ that satisfy the conditions. In this way, optimal parameters of both the primal and dual problems are found. Since more parameters are involved, an LCP is often regarded as more difficult than the primal or dual problems. However, the advantage with LCP lies with the fact that it looks at the general structure  in a given optimisation problem. This technique can sometimes make the optimisation easier. In LCP, apart from the trivial conditions arising from the constraints in primal and dual problems, there are two additional ones: the first dictated by the Lagrangian stability and the second by the complementarity slackness. These are precisely the optimality conditions shown in Eqs. (\ref{eq:medopt3}) and (\ref{eq:medopt4}). 
\\

{\it On the existence of complementary states}. Before proceeding to the geometric formulation, it is also worth mentioning how the approach of LCP has been developed. The general structure of LCP in minimum-error discrimination was first shown for discrimination between two pure qubit states in Ref. \cite{ref:hwang1}. This result was quickly generalised to mixed states \cite{ref:bae08}. These papers make full use of qubit state geometry by finding the complementary states, instead of optimising measurement. The difficulty to prove the existence of complementary states for $n$-state ($n>2$) makes it hard to generalize the results beyond the two-state case. The efforts were also made in general cases and examples as well, see Ref. \cite{ref:baejmp}.

In Ref. \cite{ref:kimura}, it was shown that the two-qubit state discrimination based on the LCP approach in Refs. \cite{ref:hwang1} and \cite{ref:bae08} can be applied to arbitrary two states in a convex operational framework in general, namely, generalised probabilistic theories such as classical and quantum theories. It is important to note the crucial step that the generalisation was possible by proving the existence of complementary states in the two-state discrimination \cite{ref:nuida}. For general cases of multiple state discrimination, the existence of complementary states was not shown and a partial result was the progress with weak Helstrom families. We also refer to specific examples of the LCP approaches for discrimination of more than two quantum states in Ref. \cite{ref:ja} where examples of three- and four-state discrimination are shown. Finally, in Ref. \cite{ref:bae11} the existence of complementary states for any set of quantum states is shown by analysing the strong duality in the semidefinite programming. This shows that the LCP approach is valid in the discrimination task in quantum theory. Finally for generalised probabilistic theories, the existence of complementarity states has been shown by proving the strong duality in convex analysis \cite{ref:baearxiv}. The LCP approach is thus valid in an convex operational framework in general. 
 \\


\noindent{\it Geometric formulation for minimum-error discrimination}. This technique essentially manipulates the optimality conditions in Eqs. (\ref{eq:medopt3})  and (\ref{eq:medopt4}) using the geometry of the state space.  We recall from the optimality conditions that the task is to search either for a symmetry operator $K$ or for complementary states $\{r_i, \sigma_i \}_{i=1}^N$.  as soon as this is done, the rest of the formulation is straightforward. To this end, we first define a polytope of given quantum states, denoted by
\bea
\mathrm{Pol} (\{q_i ,\rho_i \}_{i=1}^N ) \nonumber
\eea 
such that the vertices of the polytope correspond to state $q_i \rho_i$ for $i=1,\cdots, N$.  Let $\mathrm{Pol} (\{r_i ,\sigma_i \}_{i=1}^N )$ denote the polytope of complementary states.  By finding this polytope in the state space, complementary states can be found and optimal state discrimination can be  analysed.

Now, notice that two polytopes $ \mathrm{Pol} (\{q_i ,\rho_i \}_{i=1}^N )$ and $\mathrm{Pol} (\{r_i ,\sigma_i \}_{i=1}^N )$ of a given set of  states are congruent to each other in the state space since, from Eq. (\ref{eq:medopt3}), it follows that
\bea
q_i \rho_i - q_j \rho_j = r_j \sigma_j - r_i \sigma_i,~~\forall~i,j=1,\cdots, N, \label{eq:relpol}
\eea
showing that  (i) the lines connecting two vertices $i$ and $j$ of the respective polytopes must be equal and (ii) they are anti-parallel.  We next locate the polytope $\mathrm{Pol} (\{r_i ,\sigma_i \}_{i=1}^N )$ in the state space, so that there exists a single operator $K$ satisfying Eq. (\ref{eq:medopt3}).


\noindent{\it A general form of the guessing probability}. The optimality conditions also allow us to find a general form of the guessing probability as well as its operational meaning. Let us first write the guessing probability, given by $\tr[K]$ in Eq. (\ref{eq:geog}), as 
\bea
\tr[K] = \tr [ \frac{1}{N} \sum_{i=1}^{N} K ] = \tr[ \frac{1}{N} \sum_{i=1}^N q_i \rho_i + r_i \sigma_i ]= \frac{1}{N} + \frac{1}{N} \sum_{i=1}^N r_i. \label{eq:trK}
\eea
Note that parameters $\{ r_i\}_{i=1}^N$ correspond to the distance between the symmetry operator and corresponding states $\{ \rho_i \}_{i=1}^N$ as follows. We first recall that $K = q_i \rho_i + r_i \sigma_i$ for all $i=1,\cdots, N$ and $K \geq q_i \rho_i$, from which one can find that
\bea
r_i = \tr [ r_i \sigma_i  ]  =\tr [ K - q_i \rho_i  ] =  \| K - q_i \rho_i \|_1. \nonumber
\eea
Parameter $r_i$ in Eq. (\ref{eq:trK}) can be replaced by the relation in the above. The guessing probability is then expressed as follows,
\bea
p_{\g} = \frac{1}{N} + R(  K \|  \{ q_i ,\rho_i \}_{i=1}^N ),~\mathrm{with }~ R(  K \|  \{ q_i ,\rho_i \}_{i=1}^N ) = \frac{1}{N} \sum_i r_i \label{eq:generalform}
\eea
where $R(  K \|  \{ q_i ,\rho_i \}_{i=1}^N )$ refers to the average distance between the symmetry operation $K$ and states $\{ q_i , \rho_i \}_{i=1}^N$ \cite{ref:baenjp}. This finds the guessing probability as the probability deviated from the random guess $1/N$ by the average distance $R(  K \|  \{ q_i ,\rho_i \}_{i=1}^N )$ . 

The general form in the above corresponds to a quantum analogue of the guessing probability for classical systems. In the classical regime, random variables directly denote those alphabets in preparation and measurement while quantum states no longer play a role, see also the subsection \ref{subsec:setting}. Let $X$ and $Y$ denote random variables in preparation and measurement, respectively. Assuming that no information other than $Y$ from measurement is provided, the guessing task about $X$ can be described by the following general form \cite{ref:cs}
\bea
p_{\g} = \frac{1}{N} + d(X|Y) \nonumber
\eea
where $d(X|Y)$ is the (variational) distance of probability distribution $p_{X|Y}(x|y)$ from random $1/N$. One observes the analogy between $d(X|Y)$ and $R(  K \|  \{ q_i ,\rho_i \}_{i=1}^N )$ that both describe the variational distances from given information to the preparation in the probability-theoretic way \footnote{In quantum state discrimination, the set of quantum states in the preparation is known from the beginning. Compared to the classical case, this introduces an {\it a priori} information. The guessing probability in Eq. (\ref{eq:generalform}) is thus characterised by the distance from the {\it a priori} information, the ensemble $\{q_i ,\rho_i \}_{i=1}^N $, but not from the purely random $ I/d $.}. 

If {\it a priori} probabilities are equal, i.e. with $q_i = 1/N$ for all $i=1,\cdots,N$, there is further simplification in the form of the guessing probability. From the optimality condition in Eq. (\ref{eq:medopt3}), the guessing probability is given by $p_{\g} = \tr[K] = q_i + r_i$ for any $i=1,\cdots,N$. Since it holds for all decompositions of the operator $K$, we have that $q_i + r_i = q_j + r_j$ for all $i,j$. Since $q_i = q_j = 1/N$, we also have that $r_i = r_j$ for all $i,j=1,\cdots, N$. Thus, the guessing probability in Eq. (\ref{eq:trK}) simplifies
\bea
p_{\g} = \tr[K] =  \frac{1}{N} +r \label{eq:guessqubit}
\eea
where $r$ denotes $r_i$ for $i=1,\cdots,N$ since they are identical. 
Therefore, for the case of equal {\it a priori} probabilities, the task is reduced to finding a single parameter $r$. 


\subsubsection{ Qubit state discrimination} 
\label{subsec:qubit}

There has been much recent progress in studying the optimal methods for discrimination of qubit (2-level) states.  A recent attempt in Ref. \cite{ref:samsonov} considers pure qubit states  where the state discrimination is  expressed in terms of Bloch vectors.  Partial solutions for three- and four-qubit states have been illustrated with examples. In Ref. \cite{ref:terhal}, a systematic  analytic study of qubit state discromination has been carried out. The method exploits the dual problem, shown in Eq. (\ref{eq:dual}), in semidefinite programming. 
In Refs. \cite{ref:bae, ref:baenjp}, following a  along a similar method but with a different way of performing  semidefinite programming, an approach based on an LCP has been used to solve qubit state discrimination, see Eq. (\ref{eq:lcp}).  In the latter approach, optimal discrimination of arbitrary qubit states given at random can also be completely solved in an analytical way. Based on this approach,  a complete analysis has recently been obtained for the discrimination of a system of three qubit states with arbitrary {\it a priori} probabilities \cite{ref:hakwon}. In the following, we introduce the geometric formulation for qubit states and provide some examples. 

For qubit states, the Bloch sphere provides a convenient and a complete description with a clear geometric picture. For convenience, we write by $\rho_{i} = \rho(v_i)$ with a Bloch vector $v_i$ as follows,
\bea
\rho_ i = \rho( v_i ) = \frac{1}{2} (I + v_i \cdot \vec{\sigma})/2,  ~~\mathrm{with }~ \vec{\sigma} = (X,Y,Z)\nonumber
\eea
where Pauli matrices are denoted by $X,Y$ and $Z$. The state is pure if $\| v_i \| = 1$, otherwise, mixed. Then, as mentioned previously, the LCP deals with optimality conditions, which for qubit states $\{ q_i , \rho(v_i)\}_{i=1}^N$ are
\bea
K = q_i \rho(v_i) + r_i \sigma_i,~ ~\mathrm{and} ~~ r_i \tr [M_i \sigma_i] =0   \label{eq:comqubit}
\eea
for complementary states $\{ \sigma_i \}_{i=1}^N$, operator $K$, and parameters $r_i \geq 0$. See also Eqs. (\ref{eq:medopt3}) and (\ref{eq:medopt4}). Recall that in this approach, instead of directly finding optimal measurement, the task is to find an operator $K$ and parameters $\{r_i \geq 0\}_{i=1}^N$ and $\{ \sigma_i \}_{i=1}^N$ that satisfy the optimality conditions in the above.


Suppose that the states are assigned with equal {\it a priori} probabilities, i.e., preparation given by $\{ q_i = 1/N,\rho(v_i) \}_{i=1}^N$. The guessing probability is given by the simpler form in Eq. (\ref{eq:guessqubit}). One still needs to determine a single parameter $r$ geometrically. Recall from the subsection \ref{sec:generalapp} that two polytopes, $\mathrm{Pol}(\{ 1/N ,\rho(v_i) \}_{i=1}^N)$ with given states and $\mathrm{Pol}(\{ r ,\sigma_i \}_{i=1}^N)$ with complementary states, are congruent. Note that non-trivial complementary states must be pure states from the optimality condition in Eq. (\ref{eq:medopt4}) \footnote{This holds true for qubit states.}. The two polytopes $\mathrm{Pol}(\{ r ,\sigma_i \}_{i=1}^N)$ and $\mathrm{Pol}(\{ \sigma_i \}_{i=1}^N)$ are similar but the the ratio between the two polytopes is only indicated by $r$. Since $\mathrm{Pol}(\{ r ,\sigma_i \}_{i=1}^N)$ is congruent to $\mathrm{Pol}(\{ 1/N,\rho(v_i) \}_{i=1}^N)$, it holds that $\mathrm{Pol}(\{ 1/N ,\rho(v_i) \}_{i=1}^N)$ and $\mathrm{Pol}(\{ \sigma_i \}_{i=1}^N)$ are similar. Thus, thehe ratio $r$ is found by the relation in Eq. (\ref{eq:relpol}):
\bea
\frac{1}{N} \rho(v_i)  - \frac{1}{N} \rho(v_j)  = r (\sigma_j  - \sigma_i )~~\rightarrow~~ r = \frac{  \|   \frac{1}{N} \rho(v_i) - \frac{1}{N} \rho(v_j) \|_1    }{  \|  \sigma_i - \sigma_j \|_1  } \nonumber
\eea
where the trace norm is taken. The natural norm in the Bloch sphere is the Hilbert-Schmidt norm, which then has a property of being proportional to the trace distance by a constant:
\bea
\|\rho_1 - \rho_2 \|_1 = \sqrt{2} \| \rho_1 - \rho_2 \|_2~~~\mathrm{for~all}~\rho_1, ~ \rho_2, \nonumber
\eea
where the Hilbert-Schmidt norm is denoted by $\|\cdot \|_2$. The relation shown above means that the ratio $r$ of comparing the two lines, estimated in Hilbert-Schmidt or the trace norm, remains the same in both cases. Therefore, one can safely take either of the measures to express the ratio, $r$, by referring to the geometry of the Bloch sphere.

Note that, since $\{\sigma_i\}_{i=1}^N$ are pure states, $\mathrm{Pol}(\{ \sigma_i \}_{i=1}^N)$ is the largest polytope similar to the original one $\mathrm{Pol}(\{ 1/N ,\rho(v_i) \}_{i=1}^N)$. This means that, to find the polytope $\mathrm{Pol}(\{ r ,\sigma_i \}_{i=1}^N)$, one expands the original one $\mathrm{Pol}(\{ 1/N ,\rho(v_i) \}_{i=1}^N)$ within the Bloch sphere such that its volume is maximal in the sphere: then, some of vertices must reach the sphere. In this way, two polytopes remain similar each other and, the ratio $r$ is computed in a straightforward manner, obtaining the guessing probability directly. Optimal POVMs are obtained after complementary states are explicitly obtained, with the other optimality condition in Eq. (\ref{eq:medopt4}). This is done by rotating the expanded polytope $\mathrm{Pol}(\{\sigma_i \}_{i=1}^N)$ such that corresponding lines of the two polytopes are anti-parallel, and then the vertices of the rotated one correspond to optimal POVMs \cite{ref:bae,ref:baenjp}.

{\it Examples.} As the simplest case, let us first consider two-state discrimination. As mentioned above, this works by 
\begin{itemize}
\item constructing the polytope of given states, 
\item finding a similar polytope that is largest in the Bloch sphere, and 
\item computing the ratio between the two. 
\end{itemize}
The guessing probability is obtained, and the optimal POVMs are found accordingly by rotating the similar polytope. For two-state discrimination $\{q_i , \rho(r_i) \}_{i=1}^2$, the polytope of given states is the line that connects $q_1 r_1$ and $q_2 r_2$ in the Bloch sphere. The expanded polytope similar to the original one is the diameter parallel to the line connecting given Bloch vectors. This corresponds to the polytope of complementary states $\sigma_1$ and $\sigma_2$. Taking the trace distance, the diameter is given by $2$ and the distance of states is given by $\|q_1\rho_1 - q_2 \rho_2 \|_1$.  Since $N=2$, from Eq., we have (\ref{eq:guessqubit}) 
\bea
r = \frac{ \|  q_1 \rho_1 - q_2 \rho_2  \|_1 }{2}, ~\mathrm{and~ therefore}~~ p_{\g } =  \frac{1}{2} +\frac{1}{2}{ \|  q_1 \rho_1 - q_2 \rho_2  \|_1 }. \nonumber
\eea
giving the Helstrom bound for qubit states, see Eq. (\ref{eq:helbd}). The optimal measurement is then found by rotating the diameter so that it is anti-parallel to the original. 


\begin{figure}[t]
\begin{center}
\includegraphics[width=4in]{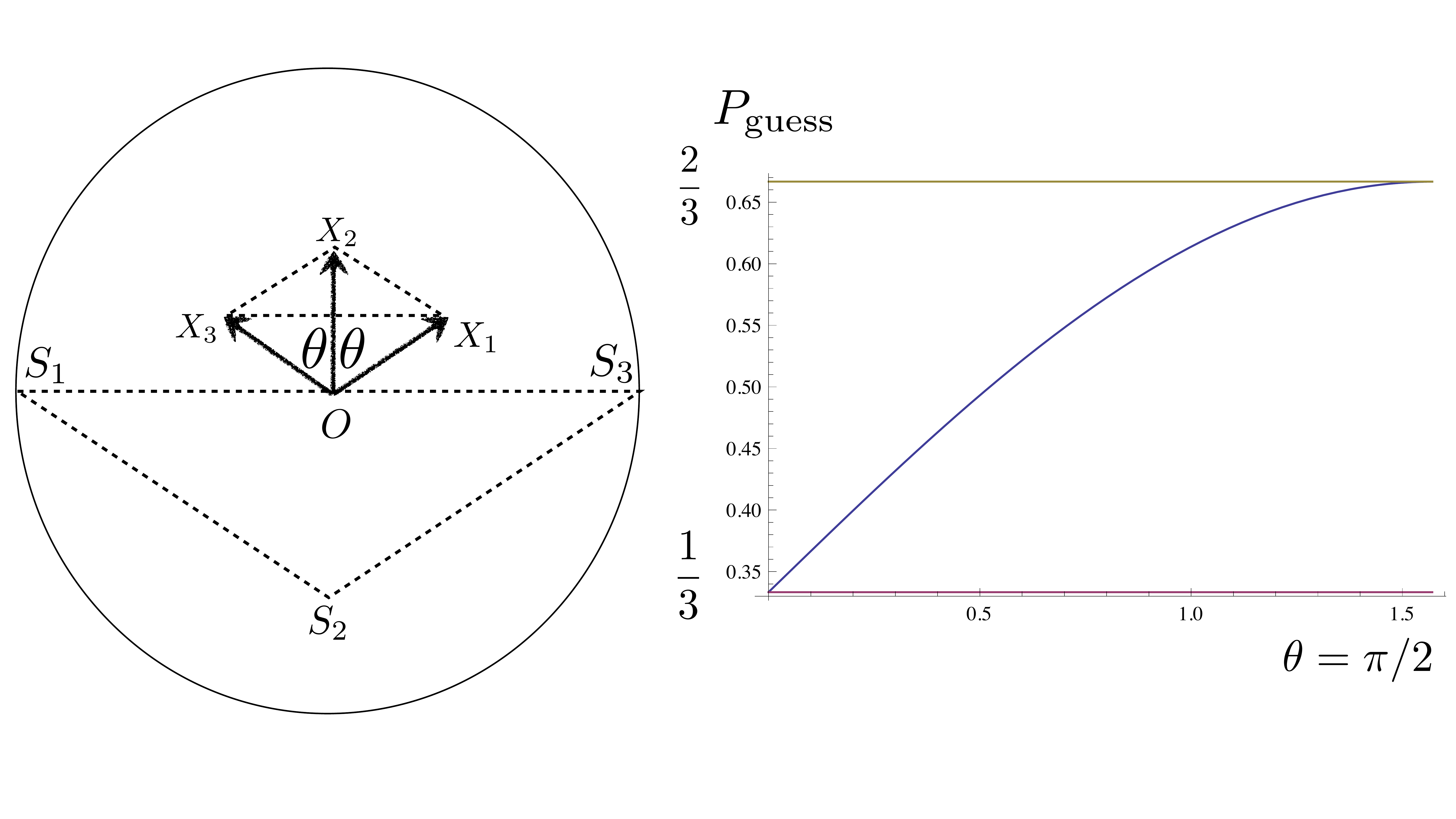}
\caption{Discrimination of three qubit states $\{q_{\x}=1/3,\rho_{\x} = |\psi_{\x}\rangle \langle \psi_{\x}| \}_{\x=1}^{3}$ lying in a half-plane is shown, see Eq. (\ref{eq:three1}). 
In the half-plane, each state $\rho_{\x}/3$ corresponds to line $OX_{\x}$ for $\x=1,2,3$. The polytope of given states corresponds to the triangle $X_1 X_2 X_3$, an isosceles triangle. The largest triangle similar to $X_1 X_2 X_3$ is given by  $S_1 S_2 S_3$. The ratio between the two triangles is, $r = (\sin\theta ) / 3$. The guessing probability is $P_{\g} = (1+\sin\theta)/3$. Note that the set $\{ OS_{\x} \} _{ \x=1,2,3 }$ corresponds to complementary states. Since the complementary state $OS_{2}$ is not pure, the optimal POVM for the state $\rho_{2}$ is the null measurement, $M_{2}=0$.} 
\label{fig:three1}
\end{center}
\end{figure}

The geometric method can be in fact applied to an arbitrary set of qubit states with equal {\it a priori} probabilities $\{ q_i = 1/N , \rho(v_i)\}_{i=1}^N$. The method may be useful  even if the given states have complex structure - see for instance polyhedron states in Ref. \cite{ref:bae} or even if there is no symmetry. Let us here consider an example of three qubit states here \cite{ref:baenjp}:
\bea
|\psi_1 (\theta)\rangle & = & \cos \frac{1}{2}(\theta_0 + \theta ) |0\rangle + \sin \frac{1}{2} (\theta_0 + \theta) |1\rangle \nonumber\\
|\psi_2 \rangle & = & \cos \frac{1}{2} \theta_0 |0\rangle + \sin \frac{1}{2}\theta_0 |1\rangle \nonumber\\
|\psi_3 (\theta)\rangle & = & \cos \frac{1}{2} (\theta_0 - \theta ) |0\rangle + \sin \frac{1}{2} (\theta_0 - \theta) |1\rangle, \label{eq:three1} 
\eea
forming an isosceles triangle in the Bloch sphere. In Fig. (\ref{fig:three1}), the three states are depicted in the Bloch sphere. Fig(\label{fig:three1}) also shows how the guessing probability and the optimal POVMs can be obtained geometrically. The guessing probability for $\theta \leq \pi/2$ is 
\bea
p_{\g} = \frac{1}{3} (1 + \sin 2\theta). \nonumber
\eea
For $\theta \geq \pi/2$, the ratio $r$ is obtained as $1/3$, and the guessing probability is given by 
\bea
p_{\g} = 1/3 + 1/3 = 2/3.\nonumber
\eea 
Note that an interesting observation here that the guessing probability remains the same in ranges $\theta\leq \pi/2$ or $\theta\geq \pi/2$.  The result is the same as in Ref. \cite{ref:jozsa}, where the  von Neumann entropy is taken  as a measure for the distinguishability. Note that the guessing probability is also an entropic measure, called the min-entropy \footnote{The guessing probability corresponds to the conditional min-entropy, $H_{\min}  = -\log p_{\g}$. See Sec. \ref{sec:min-entropy}}. From the two measures, the agreement shows that distinguishability of quantum states is a global property that cannot be reduced to properties of pairs of states. \\


Using the geometric method, the optimal discrimination of geometrically uniform qubit states, see also the subsection \ref{subsec:multi}, is easily reproduced. Let us consider a simple case of geometrically uniform states $\{1/3, \rho_i = |\psi_i\rangle \langle \psi_i | \}$ where
\bea
|\psi_1 \rangle  & = & \cos\theta_0 |0\rangle +e^{2\pi/3} \sin\theta_0 |1\rangle  \nonumber \\
|\psi_2 \rangle  & = & \cos\theta_0 |0\rangle + \sin\theta_0 |1\rangle  \nonumber \\
|\psi_3 \rangle  & = & \cos\theta_0 |0\rangle +e^{-2\pi/3} \sin\theta_0 |1\rangle.  \label{eq:3geo}
\eea
These are instances of the set in Eq. (\ref{eq:three1}) for $\theta > \pi/2$. Based on the geometric method as illustrated in Fig. (\ref{fig:three1}), the guessing probability is found to be  $p_{\g} = 2/3$. The optimal POVMs are found by rotating the polytope: they are explicitly given by $\{M_i = |\phi_i \rangle \langle \phi_i | /3 \}_{i=1}^3$ where
\bea
|\phi_1\rangle & = &  (|0\rangle + |1\rangle) / \sqrt{2}  \nonumber \\
|\phi_2 \rangle & = &   (|0\rangle + e^{2\pi i /3} |1\rangle)/ \sqrt{2} \nonumber \\
|\phi_3 \rangle & = & (|0\rangle + e^{- 2\pi i /3} |1\rangle)/ \sqrt{2}.\label{eq:3geom}
\eea

\subsection{Selected topics in applications of two-state discrimination}

Indeed, two-state discrimination is the only case where the optimal discrimination is completely analysed. In this subsection, we select and consider a few applications of two-state discrimination. It quantifies state preparation, classifies the measurement strategies, and introduces norms  that possess operational meanings.


\subsubsection{Quantification of quantum state preparation}  Suppose that, for a certain task, the goal is to prepare a system to be in an ideal state $\rho_{\mathrm{ideal}}$, denoted by $S_{\mathrm{ideal}}$. Often, the prepared state is not the original state but a slghtly different state $\rho_{\mathrm{actual}}$. This state can be identified through quantum state tomography. Note also that two different states (ideal vs actually prepared) do not always give different measurement outcomes.   An operational way that can quantify the two systems $S_\mathrm{ideal}$ and $S_\mathrm{actual}$ is to say that they are the same if they cannot be distinguished by any measurements.   The guessing probability for the two states is
\bea
p_{\g} (S_{\mathrm{ideal}},S_{\mathrm{actual}}) = \frac{1}{2} + \frac{1}{2} \| \frac{1}{2} \rho_{\mathrm{actual}} - \frac{1}{2} \rho_{\mathrm{real}} \| \nonumber 
\eea
and it can be used to distinguish the prepared state $\rho_{\mathrm{actual}}$ to the desired one $\rho_{\mathrm{ideal}}$.  This way of quantification via two-state distinguishability is used in a lot of contexts in quantum information applications involving state preparation. A good example is quantum key distribution where the desired state is a maximally entangled state needed for generating an ideal key, and security with an actual state obtained after key distillation techniques is quantified by the distinguishability of the two different states \cite{ref:rennerthesis}. In addition,  two-state distinguishability is also an important tool to quantify security of building blocks of large cryptographic systems, for universally composable security \cite{ref:rennerthesis, ref:ran} in quantum cryptography. \\

\subsubsection{Distinguishing measurements}  

Let us first recall two-state discrimination $\{q_i , \rho_i \}_{i=1}^2$ for which the guessing probability could be written as
\bea
p_{\g} = \frac{1}{2} + \max_{-I \leq M \leq I } \tr[M X ],~\mathrm{with}~ X = q_1 \rho_1 - q_2 \rho_2.  \nonumber  
\eea
One finds that distinguishing two quantum states is completely characterised by the operator $M$, see Eq. (\ref{eq:optm2}).  If the two quantum states are fixed as well as operator $X$, there can be various ways of choosing measurements giving rise to different upper bounds for the guessing probability.  This perspective is related to the process of classifying measurements, especially for multipartite quantum systems. 
They are, namely, non-local, separable, and local measurements, or local operations and classical communication (LOCC). In fact, it turns out that that the set of separable operations is strictly larger than the set of LOCC processes \cite{ref:seplocc}. Since it is obvious that separable operations are a class of quantum operations, it follows that 
\bea
\mathrm{LOCC} \subsetneq  \mathrm{SEP} \subsetneq \mathrm{ALL}. \nonumber
\eea
The strict inclusion of LOCC to SEP is shown using quantum state discrimination based on the set of quantum states that can be perfectly distinguished by separable operations but not by LOCC protocols \cite{ref:seplocc}. 

In general, the characterisation of the capabilities of LOCC remains an open problem, see, for instance, the recent progress in Ref. \cite{ref:all-locc}. This characterisation is important since LOCC is the alternative way to study entangled states that are resources for quantum information processing in general. At the same time, it is also closely related to the separability problem that underlines the border between entangled and separable states.  \\


\subsubsection{ Norm defined by operational meanings} Motivated by the expression in Eq. (\ref{eq:guess2}), the trace-norm in Eq. (\ref{eq:1norm}) has been generalised to a (semi-) norm characterised by the choice of measurement $M$. Since the trace-norm possesses an operational meaning in terms of distinguishability via general quantum measurement, a new norm is introduced based on possible quantum measurements.  To this end, it is useful to express the trace norm as 
\bea
\| X \|_1 = \max_A \tr [AX]  = \max_{\{ A_x \geq 0,\sum_{x} A_{x} = I \}} \sum_{x} |  \tr A_x X |.   \nonumber
\eea
A norm induced by measurement $M$ can then be constructed by restricting the possible measurements on $M$ as follows
\bea
\| X \|_{M} = \sup_{M } \tr[MX]  = \sup_{M= \{M_x \}_{x=1}^N: \mathrm{POVM}} \sum_{x} | \tr X M_x  |.  \nonumber
\eea
where $M$ can be chosen as a class of measurement: LOCC, separable measurement etc. \cite{ref:winterM}. The trace-norm is recovered by setting $M$ as the most general measurement strategy. The formalism is useful for describing the gaps between different classes of quantum operations, for instance, between $ \| X \|_{\mathrm{SEP}} $ and $\| X \|_{\mathrm{LOCC}} $. See also the recent and related progress along the line and its connections to quantum designs \cite{ref:lancien}. 


\subsection{ Unambiguous state discrimination }

\begin{figure}[t]
\begin{center}
\includegraphics[width=5.5in]{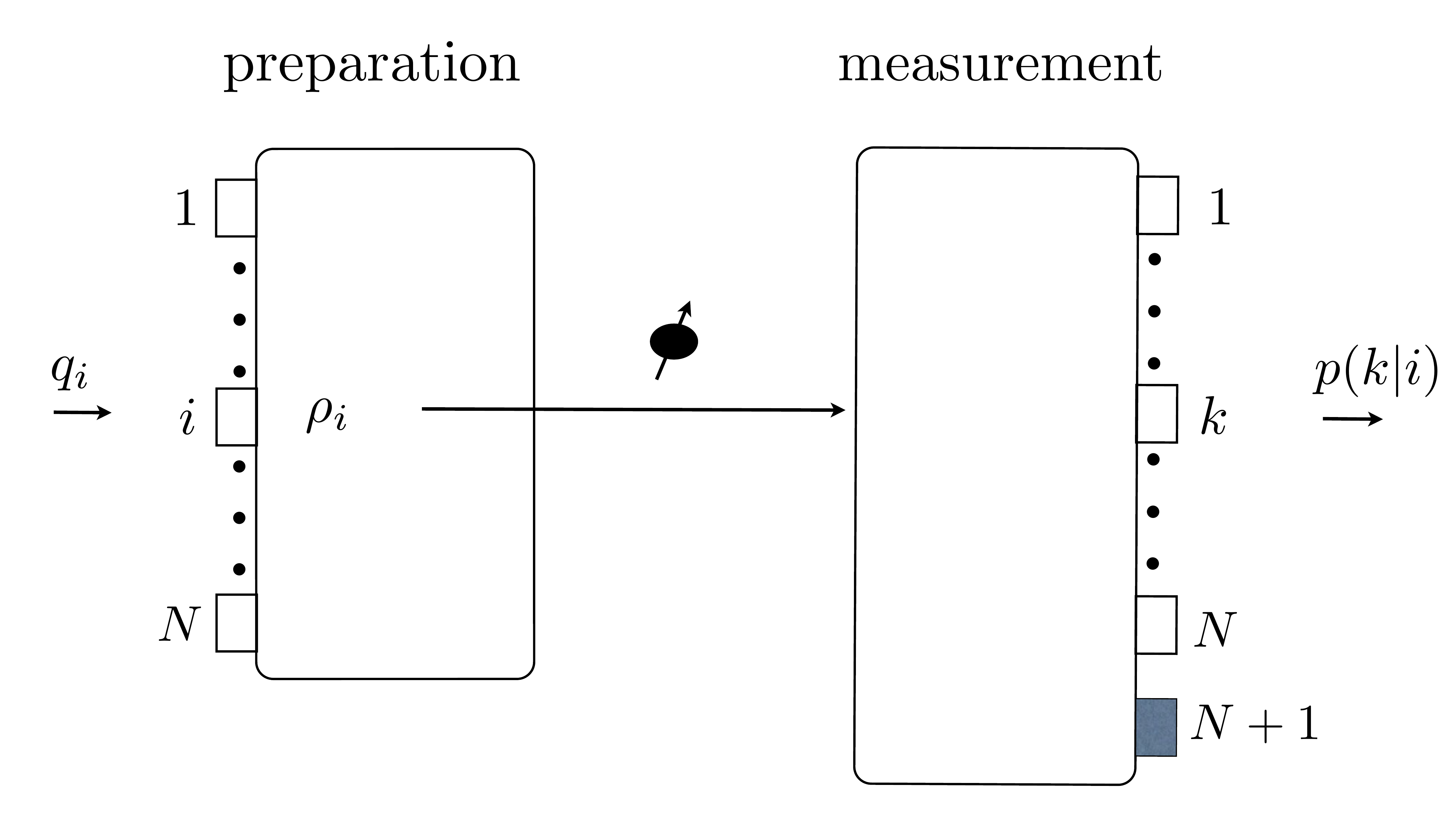}
\caption{A setting for unambiguous discrimination is shown. There is an additional output port $N+1$ that collects all ambiguous results; detection events on this port do not make any conclusion about which state has been detected. Detection events on other output ports have no ambiguity; a detection event on the $i$th port means that $\rho_i$ is detected with certainty.   }
\label{fig4:fig4}
\end{center}
\end{figure}

In minimum-error state discrimination, the measured state in the output ports  does not necessarily correspond to the prepared state. There is always an element of error.  In  unambiguous discrimination,  for a given set of  states $\{ q_i ,\rho_i\}_{i=1}^{N}$, there exists measurements $\{ M_{i} \}_{i=1}^{N+1}$ with an additional element $M_{N+1}$ such that
\bea
p(i | j) = \tr[M_i \rho_j ] =0,~\mathrm{for} ~ i,  j = 1,\cdots, N. \nonumber
\eea
This condition ensures that detection event on $M_k$ is associated only with the state $\rho_k$ and not with others.  However, there is an additional outcome, corresponding to $M_{N+1}$, that gather all the ambiguous results, see Fig. (\ref{fig4:fig4}). Note that the additional POVM $M_{N+1}$  should fulfill the completeness condition, i.e., in case that $\sum_{i=1}^N M_i \neq  I$, one can find an additional POVM such that $\sum_{i =1 }^{N+1} M_i =I$. This is not always possible for arbitrary states. For pure states, unambiguous discrimination can be designed if they are linearly independent \cite{ref:chefles}. For two mixed states, this is done when the support for the states do not completely overlap \cite{ref:spekkens}. 

For the given states $\{ q_i, \rho_i\}_{i=1}^N $, unambiguous state discrimination is described by
\bea
\max && \sum_{i=1}^{N} q_i \tr[M_i \rho_i] \nonumber\\
\mathrm{subject ~ to} &&  \tr[M_i \rho_{j}] =0~ ~\mathrm{for} ~i \neq j = 1,\cdots, N \nonumber\\
&&  \sum_{i=1}^{N+1} M_i = I  ~~\mathrm{and} ~~ M_i\geq 0~\mathrm{for} ~i=1,\cdots,N+1.  \nonumber
\eea
A general numerical method for solving unambiguous state discrimination is to apply semidefinite programming \cite{ref:eldarsdp2}. The error comes only from the collection of ambiguous results, and thus it suffices to minimise the probability $\tr[M_{N+1} \rho]$ where $\rho = \sum_{i}q_i \rho_i$. In general, unambiguous state discrimination is not always possible for linearly dependent quantum states \cite{ref:chefles}. For instance, for qubit states, unambiguous discrimination is possible only when the given states are two pure states. For two states $|\psi_1 \rangle$ and $|\psi_2\rangle$ with {\it a priori} probabilities $1/2$, the probability of having ambiguous results is
\bea
P_{IDP} = | \langle \psi_1 | \psi_2 \rangle |. \nonumber
\eea
The last result depends on the overlap between the two states. This probability   is called the Ivanovic-Dieks-Peres limit \cite{ref:iva, ref:die, ref:per}. 

If there are more than two pure states, a recent approach discussed in Ref. \cite{ref:bergou3} is used. This approach uses the geometrical properties of the given states, and it can  be applied to arbitrary number of linearly independent pure states. Remarkably, for linearly independent three pure states, unambiguous state discrimination has been obtained in a closed form. 

Unambiguous state discrimination has a number of applications: in particular, for state comparison and state filtering. It is also connected to the entanglement concentration. For more details, the readers may wish to refer to the excellent reviews in Refs. \cite{ref:rev1, ref:rev2, ref:rev3}.


\subsection{Maximum confidence discrimination  }

In unambiguous state discrimination, a detection event inevitably corresponds to a correct guess. This is in contrast to the minimum-error state discrimination where a detection event on an output port does not lead to a correct guess all the time. However, unambiguous discrimination can be applied to limited cases, for instance, when given quantum states are linearly independent or when the states do not have an identical support. 

In Ref. \cite{ref:cro}, maximum confidence measurement is introduced as an independent discrimination strategy that utilises the confidence on detection events as the figure of merit. This method coincides with unambiguous state discrimination if the latter technique applies. Or, it corresponds to minimum-error discrimination if the confidence is maximised on the average over all given states. 
Here, the confidence is meant by the probability that a detection event which has happened in an output port leads to giving a correct guess. The maximum confidence measurement is an approach of retrodictive quantum theory that maximises retrodiction from the measurement rather than the prediction from the preparation. 

In maximum confidence discrimination, the optimal measurement is obtained when a detection event on each port provides a correct answer with highest probability. Otherwise, there is an extra output port $N+1$ that collects all ambiguous answers. More precisely, for states $\{q_i, \rho_i \}_{i=1}^{N}$, suppose one obtains the measurement outcome on an output port $M_k$, indicating that state $\rho_k$ has been detected. The probability that  the state $\rho_k$ has actually occurred, given that a detection event is shown at $M_k$, can be analyzed using Bayes' rule \cite{ref:cro},
\bea
p( \rho_k |M_k) = \frac{ p(\rho_k) p( M_k | \rho_k)  }{ p(M_k)}  = \frac{q_k \tr[M_k \rho_k]  }{\tr[ \rho M_k ]},~\mathrm{with~}\rho = \sum_i q_i\rho_i ,  \label{eq:conpro}
\eea
where $p(M_k)$ denotes the probability that measurement port,  $M_{k}$, is clicked, and $p(\rho_k)$ the probability that the state $\rho_k$ is prepared from one of states in the ensemble, $\rho = \sum_i q_i \rho_i$. The maximum confidence discrimination is then obtained by maximising the conditional probabilities 
\bea
\max && p( \rho_k |M_k)  ~~\mathrm{for}~k=1,\cdots,N\nonumber \\
\mathrm{subject ~to} && \sum_{i=1}^{N+1} M_i = I,~~ M_i \geq 0~\forall~i=1,\cdots,N+1 \nonumber
\eea
where measurement outcomes in the additional POVM $M_{N+1}$ corresponds to ambiguous results. Note that the additional POVM element $M_{N+1}$ is essential since optimal POVM elements $\{ M_i\}_{i=1}^N$ do not form a complement measurement in general. 

\subsubsection{Example}

A simple and useful example that compares the maximum confidence measurement with the minimum-error state discrimination is provided in Ref. \cite{ref:cro} using the geometrically uniform three-qubit states: $\{ 1/3, \rho_i = |\psi_i \rangle \langle \psi_i|  \}_{i=1}^3 $. These states are linearly dependent and thus unambiguous discrimination is not possible. The optimal measurement for the maximum confidence is given by the POVM  $ M_i = \alpha_i | \phi_i \rangle \langle \phi_i  |$ for $i=1,2,3$ and $M_4$ for inconclusive results, where $\alpha_1 = \alpha_2 = \alpha_3 = (2\cos^2\theta)^{-1} $ and 
\bea
|\phi_1 \rangle &=& \sin\theta |0\rangle + \cos\theta|1\rangle \nonumber \\
|\phi_2 \rangle &=& \sin\theta |0\rangle + e^{2\pi i/3}\cos\theta|1\rangle \nonumber \\
|\phi_3 \rangle &= & \sin\theta |0\rangle + e^{-2\pi i/3} \cos\theta|1\rangle \nonumber  \\
M_4 & = & (1 - \tan^2\theta) |0\rangle \langle 0|. \nonumber
\eea
Compared to the optimal measurement for minimum-error state discrimination in Eqs. (\ref{eq:3geo}) and (\ref{eq:3geom}), optimal measurement for maximum statistical confidence depends on $\theta$. With the optimal measurement, the conditional probability in Eq. (\ref{eq:conpro}) is given by 
\bea
[p( \rho_i | M_i )  ]_{\max } = \frac{2}{3}~~\mathrm{for~all~}i=1,2,3. \nonumber
\eea
whereas, with optimal measurement for the minimum-error state discrimination, the conditional probablity is given by $(1+ \sin2\theta)/3$. 

\subsubsection{Optimal measurement}

In this section, we briefly review the general method of finding optimal measurement for maximum confidence measurement. For given states $\{ q_i , \rho_i \}_{i=1}^N$, the main technique is to consider an ansatz
\bea
M_i = c_i \rho^{-1/2} Q_i \rho^{-1/2},~~ \mathrm{where}~~\rho = \sum_{i} q_i \rho_i \label{eq:mcm}
\eea
with  constant $c_i$, that needs to be determined, and an operator $Q_i$. By substituting this expression into the conditional probability in Eq. (\ref{eq:conpro}), we have
\bea
p(\rho_k |M_k) = q_k \tr[\rho_k \rho^{-1}] \tr[ \rho_{k}^{'} Q_k]~~\mathrm{where}~ \rho_{k}^{'} = \frac{\rho^{-1/2} \rho_k \rho^{-1/2} }{\tr[\rho_k \rho^{-1} ]}   \nonumber
\eea
To maximise the conditional probability, the operator $Q_k$ should be taken as an operator having a maximal overlap with the state $\rho_{k}^{'}$.  We denote the spectral decomposition of the state by $\rho_{k}^{'} = \sum_{i} \lambda_{i}^{k}  | \lambda_{i}^{k}\rangle \langle \lambda_{i}^{k} |$ and let $\lambda_{\max}^{k} = \max_{i} \lambda_{i}^{k}$.  We then have $Q_k = | \lambda_{\max}^{k} \rangle \langle \lambda_{\max}^{k}|$. This last step determines  optimal POVMs. The coefficients $\{ c_i\}_{i=1}^N$ are then determined under the constraint that measurement is complete.

\subsubsection{Equivalence to minimum-error state discrimination}

As alluded before, for individual states, the conditional probability obtained for maximum confidence measurement is always larger than that from minimum-error discrimination. One then asked if the maximum confidence measurement would attain minimum-error discrimination when maximum confidence measurement is averaged over all states. This is indeed the case since
\bea
\max \sum_{i} p(M_i) p(\rho_i | M_i )  = \max \sum_{i} p(M_i) \frac{p(\rho_i) p( M_i | \rho_i )  }{p(M_i)} =   \max\sum_i  p(\rho_i) p(M_i | \rho_i ).  \nonumber
\eea
The left-hand-side shows the maximisation of the average conditional probabilities over all measurement and it turns out to be equal to the right-hand-side, i.e. the guessing probability. Thus, it holds that maximising conditional probabilities on average is equivalent to finding the guessing probability.

\subsection{Relation between different strategies of state discrimination  }

There have been approaches that generalise both methods of minimum-error discrimination and unambiguous discrimination. Maximum confidence measurement can also be viewed as a generalisation of these approaches, in the sense that it coincides with unambiguous discrimination for linearly independent states \cite{ref:chefles} or states whose support do not overlap \cite{ref:spekkens} and with minimum-error discrimination if the average confidence is maximised \cite{ref:rev4}. These connections are obtained even though maximum confidence measurement is originally formulated for maximising the confidence of the guessing task on detection events. Maximum confidence measurement therefore interpolate between the two discrimination techniques in a quantitative way by relating an error rate and a rate of inclusive outcomes, motivated through practical conditions in experiment where errors are sometimes inevitable \cite{ref:steinberg}. 

One the one hand, minimum-error discrimination can be generalised with an additional measurement output port that collects inconclusive outcomes. By fixing the rate of inconclusive outcomes, the optimisation is achieved through the minimisation of  the error-rate. For instance, in two-state discrimination $\{q_i , \rho_i \}_{i=1}^2$, measurement is performed by three POVM elements $\{ M_1 M_2, M_3\}$ with the additional $M_3$, leaving the rate of inconclusive outcomes $Q  = \tr[\rho M_3]$ for the ensemble $\rho= \sum_{i=1}^2 q_i \rho_i$ fixed. The POVMs $M_1$ and $M_2$ are then optimised to minimise the error probability $p_{\mathrm{error}}$. The case when $Q=0$ corresponds to minimum-error discrimination. If given states are linearly independent, unambiguous discrimination can be recovered by putting $Q$ sufficiently high. This was suggested and studied in Ref. \cite{ref:int1} and further analysed in semidefinite programming with examples in Ref. \cite{ref:inteldar}. 

On the other hand, a generalisation can also be obtained by introducing an error probability to unambiguous discrimination $p_{\mathrm{error}}$. The task then is to minimise the rate of inconclusive outcomes $Q$ subject to a constraint on the error probability $p_{\mathrm{error}} \leq p_c$ for some $p_c\in[0,1]$. Unambiguous discrimination is found by putting $p_c = 0$. This method of state discrimination has been considered in Ref. \cite{ref:steinberg} and analysed for two-state discrimination in Refs. \cite{ref:horibe1, ref:horibe2}. 

In fact, relations between the two approaches is found as a trade-off between the error rate $p_{\mathrm{error}}$ and the rate of inconclusive outcomes $Q$. Recently, the relation $p_{\mathrm{error}} (Q)$ or equivalently $Q(p_{\mathrm{error} })$ is obtained analytically \cite{ref:intbagan, ref:intherzog}. 
Although the relation is not yet fully analysed for arbitrary set of quantum states, it is remarkable that some general results for the problem have been obtained, for instance optimality conditions, useful for deriving analytic results for certain symmetric cases, and a general approach for $p_{\mathrm{error}}(Q)$ with fixed $Q$, see details in Refs. \cite{ref:intbagan, ref:intherzog} and references therein. 

Finally, let us summarise a general way of transforming state discrimination with fixed rate $Q$ to a problem of minimum-error discrimination that minimises the error rate $p_{\mathrm{error}}$ \cite{ref:intbagan}. For states $\{ q_i , \rho_i \}_{i=1}^N$, let $\{ M_{i} \}_{i=1}^{N+1}$ denote POVM elements. Suppose that $M_{N+1}$ is fixed, and thus $Q = \tr[\rho M_{N+1} ]$ where $\rho = \sum_{i}q_i \rho_i$. The key idea is to identify the support of measurement comprised only of $\{ M_i \}_{i=1}^N$. Denoted by $\Omega = \sum_{i=1}^N M_i = I - M_{N+1}$, one then find POVM elements in an alternative form as,
\bea
\widetilde{M}_{i} = \Omega^{-1/2} M_i \Omega^{-1/2},~~\mathrm{for}~ i = 1,\cdots, N. \label{eq:intm}
\eea
Projecting the given states on the support, we get normalised states $\{ \widetilde{\rho}_i  \}_{i=1}^N $ and $a~priori$ probabilities $\{\tilde{q}_i \}_{i=1}^N$,
\bea
\widetilde{\rho}_{i} = \frac{  \Omega^{1/2} \rho_i \Omega^{1/2} }{\tr[\Omega \rho_i ]},~~\mathrm{and}~~\widetilde{q}_i = \frac{q_i \tr[Q \rho_i]}{1-Q}. \nonumber
\eea
The guessing probability is then found by minimum-error discrimination of $\{ \widetilde{q}_i ,\widetilde{\rho}_i \}_{i=1}^N$. It is interesting to observe that optimal strategies in Eq. (\ref{eq:intm}) share similarity with maximum confidence measurement in Eq. (\ref{eq:mcm}).

\subsection{Maximising mutual information }

In terms of communication capacity, one can also consider the maximisation of  the mutual information shared between Alice (who prepares the state) and Bob (who applies the measurement). The mutual information is used to estimate the rate of information transmitted in communication channels. For the preparation $\{q_i,\rho_i \}_{i=1}$,  the mutual information is given by
\bea
I(A:B) = \sum_i q_i  \sum_{j} p_{B|A} (j|i) \log \frac{p_{B|A} (j|i ) }{ \sum_{k} q_k p_{B|A} (j| k)} \nonumber
\eea
where $p(j |i)$ denotes the conditional probability that Bob has outcome $j$ when Alice sends state $\rho_i$ i.e. $p(j |i) =\tr[M_j \rho_i]$. Once the mutual information is maximised, the quantity is called accessible information, also known as Holevo information \cite{ref:holevochi},
\bea
I(A:B) \leq \chi = S (\rho) - \sum_i p_i S(\rho_i)
\eea
where $\rho$ denotes the ensemble and $S$ von Neumann entropy. It quantifies Bob's information accessible to Alice's through measurement on his quantum states. 

Although the upper bound has been known, it is not straightforward to find the optimal strategy. An important property is that the above mutual information is a convex function of Bob's  POVMs.  More precisely, suppose that $M^{(B)} $ is a convex combination of of measurements $M_{i}^{(B)}$, i.e. $M^{(B)} = p_1 M_{1}^{(B)} + \cdots + p_n M_{n}^{(B)} $, then it follows that
\bea
I(A:B) \leq \sum_i p_i  I(A:B_i ) \leq \max_i I (A:B_i). \nonumber
\eea
This implies that the optimisation problem can be solved by convex optimisation techniques and has a unique solution. 

As an example, in Ref. \cite{ref:sas}, geometrically uniform states in two-dimensional Hilbert space  (qubit states) are considered. For $N$-state discrimination, the symmetry is given by the following unitary transformation, a rotation about $y$-axis, and the states are obtained by successive applications to a fixed state $|\psi_0\rangle$ as follows:
\begin{displaymath}
V = \exp ( - i \frac{\pi}{N}Y) = 
\left( \begin{array}{ccc}
\cos\frac{\pi}{N} & -\sin\frac{\pi}{N}  \\
\sin\frac{\pi}{N} & \cos\frac{\pi}{N}  
\end{array} \right), ~~\mathrm{with}~~
|\psi_{0} \rangle = \left( \begin{array}{ccc}
1  \\
0
\end{array} \right),~ \mathrm{and}~|\psi_k \rangle = V^k |\psi_0\rangle.
\end{displaymath}
Using the symmetry and group-theoretical properties, optimal measurement could be obtained as POVMs $\{ M_k \}_{k=0}^{N-1}$ where $M_{k}= | a_k \rangle \langle a_k | /N$, with $|a_0\rangle$ orthogonal to $|\psi_{0}\rangle$. In this case, applying the techniques in Ref. \cite{ref:bae}, this set of POVMs coincides with the optimal measurement for minimum-error discrimination of the same qubit states, even though optimal measurement for minimum-error is not unique in general. 

Note that channel capacities mentioned above are in fact defined in terms of mutual information. In this case, given a channel, the task is to optimise the mutual information over all possible sets of quantum states. For the progress along the line, there is a recent review Ref. \cite{ref:reviewgio}.
 
\subsection{  State discrimination in the asymptotic limit}
\label{subsec:limit}

Until now, quantum state discrimination is considered in a single-shot manner, where quantum states are prepared and collected,  and measurement is applied only once for the state discrimination. However, in situations where it is possible to repeat measurements, a more general setting allowing for different strategies of the measurements is possible. Instead of the figures of merit introduced in single-shot discrimination scenarios, for instance, the guessing probability, the focus is now on the statistical behaviour of the success or error probability, or extending different measurement strategies depending on costs of measurements or the inherent experimental requirements. 


\subsubsection{ Minimum-error discrimination of $i.i.d.$ states }

Suppose that, during the preparation stage, a quantum state (either $\rho_1$ or $\rho_2$) is repeatedly and independently generated $n$ times. From Bob's perspective, they are prepared in the form of $\rho_{i}^{\otimes n}$ for $i=1,2$, which are independently, identically, and distributed (i.i.d) states. The analysis on discrimination of the $i.i.d.$ states is technically hard but in fact crucial to quantum communication or quantum coding. Here, let us briefly review quantum Chernoff bound, recently obtained, that provides a complete analysis on the statistical behaviour. 

For the case of two-state discrimination between $\rho_{1}^{\otimes n}$ or $\rho_{2}^{\otimes n}$, the optimal discrimination has been completely analysed, as it is shown the Helstrom bound in Eq. (\ref{eq:helbd}). It is however not straightforward to compute the trace distance between the two states $\rho_{1}^{\otimes n}$ or $\rho_{2}^{\otimes n}$ in an analytic form. In the asymptotic limit where the error tends to be very small, the convergence rate can be easily treated by introducing the critical exponent of the convergence, as follows
\bea
p_{\mathrm{error,n}} = 1- p_{\g} (\rho_{1}^{\otimes n}, \rho_{2}^{\otimes n})  \approx \exp (-n \xi_{QCB}) \nonumber
\eea
where $p_{\g} (\rho_{1}^{\otimes n}, \rho_{2}^{\otimes n})$ denotes the guessing probability assuming equal {\it a priori} probabilities, and $\xi_{QCB}$ is called quantum Chernoff bound. 

It is also worth to note that for the classical analogy, i.e. for probabilistic systems described by $p_1$ and $p_2$, the critical exponent  is given by
\bea
\xi_{CB} = - \log (\min_{0\leq s \leq 1 }  \sum_{i} (p_{0}(i))^s (p_{1}(i))^{1-s}  ). \nonumber
\eea
Recently in Ref. \cite{ref:aude2007}, its explicit form has been derived as a precise analogy to its classical counterpart:
\bea
\xi_{QCB} = \lim_{n \rightarrow \infty}  -\frac{p_{\mathrm{error},n} }{n} = - \log (\min_{0\leq s \leq 1} \tr [~\rho_1^{s} \rho_{2}^{1-s}]  ). \label{eq:qcb}
\eea

The convergence rate of errors in quantum state discrimination has been extended to multiple states.  Let the given states be given by a set $\Sigma = \{ \rho_i \}_{i=1}^N$ and let $\xi_{QCB} (\Sigma)$ denote quantum Chernoff bound for the set $\Sigma$:
\bea
\xi_{QCB} (\Sigma )= \min_{i,j} \xi_{QCB} (\rho_i , \rho_j)~~\mathrm{for}~i,j=1,\cdots, N. \nonumber
\eea
This expression involves the pairwise quantum Chernoff bound among all pairs. It has been conjectured that it suffices to consider just a pair of two states having a minimal overlap. This holds true if the given states are commuting so that they can be considered as classical states. Otherwise, it is known that the quantum Chernoff bound for multiple states has the following upper and lower bounds:
\bea
\frac{1}{3} \xi_{QCB } (\Sigma)\leq \lim_{n\rightarrow \infty} -\frac{p_{\mathrm{error,n}} (\Sigma) }{n} \leq \xi_{QCB} (\Sigma),
\eea
where $p_{\mathrm{error,n}} (\Sigma)$ denotes the error probability for the states in the set $\Sigma$ assuming equal {\it a priori} probabilities. It remains an open question whether the factor $1/3$ in the lower bound can be removed. A conjecture is that the quantum Chernoff bound for multiple states is equal to the Chernoff bound of two states having the maximal overlapping. See also the recent progress along the line \cite{ref:aude2008, ref:nuss2009, ref:nuss2011}.

\subsubsection{Measurement strategies}

For state discrimination among $i.i.d.$ states $\Sigma = \{ \rho_i^{\otimes n}\}_{i=1}^N$, the optimal measurement is the so-called ``collective measurement", which is the most general form of measurements on multiple copies of states. In this scheme, each POVM element $M_j^{(n)}$ for $j=1,\cdots,N$ on $n$ copies works on the whole block collectively as follows 
\bea
p(j | i ) = \tr[M_j^{(n)} \rho_i^{\otimes n}]. \nonumber
\eea
To implement collective measurement, there is however an experimental requirement that one should be able to store quantum states for sufficiently long time, perhaps in a device called a quantum memory.  Realising quantum memory however is still experimentally challenging.  In addition, one should be able to devise a method of realizing complicated interactions among copies. This is also highly non-trivial with the present-day technology. Therefore, from a practical point of view, the implementation of collective measurement is not currently feasible. 

A possible scheme in current experiment scheme is to apply measurement on single copies individually, hence, called individual measurement. One of the most elementary strategies of this class is that an identical measurement is repeatedly applied, called repeated measurement. That is, a POVM element is of the following form: $M_j^{(n)} = m_{j}^{\otimes n} $, where $m_j$ denotes a POVM on single copies, so that 
\bea
\tr[M_j^{(n)} \rho_{i}^{\otimes n}]  = \tr[ m_j \rho_i] \times   \cdots \times \tr[ m_j \rho_i]. \nonumber
\eea
There is an even finer scheme, further elaborated, called adaptive measurement where measurement outcomes of an individual copy are updated and guides the measurement for the next copy. In this case, a POVM element applying single-copy measurement can be expressed as,
\bea
M_j = m_{j,n} [x_{j,n-1}]  \otimes \cdots  \otimes m_{j,k}[x_{j,k-1}]  \otimes  \cdots \otimes  m_{j,1}  \label{eq:indm}
\eea
where $x_{j,k}$ denotes the measurement outcome of $m_{j,k}$. This can be regarded as Bayesian updating. In the most general strategy of individual measurement, the $n$-th measurement may depend on the previous $n-1$ outcomes. In this case, a POVM would be of the form of Eq. (\ref{eq:indm}) with single-copy measurement $m_{j,k} = m_{j,k} [ \vec{x}_{j,k} ] $ where $\vec{x}_{j,k} = (x_{j,1},\cdots, x_{j,k-1} )$ the collection of the previous $k-1$ measurement outcomes $x_{j,m}$ for $m=1,\cdots,k-1$.

It remains an open question whether individual measurement is as good as collective ones, or if not, at least in the asymptotic limit \footnote{This is the 31st problem in the list in http://qig.itp.uni-hannover.de/qiproblems/.}. It is also a question on the power of quantum memory in improving distinguishability of quantum states \cite{ref:rennermemory}. In Ref. \cite{ref:twocol}, two-state discrimination on multiple copies is in fact the case. 


\section{ State Discrimination as a Tool }
\label{sec:tool}

One of important contributions of quantum state discrimination is that it provides a link between quantum foundations and quantum information processing. This link has provided much motivation to investigate fundamental problems in quantum foundations.  For instance, minimum-error state discrimination has been applied to certifying the dimension of quantum systems used in state preparation. It is an important application in the context of device-independent quantum communication where the assumptions on devices are relaxed so that a higher level of security is attained. Or, it has also been shown that optimal state discrimination is equivalent to optimal quantum cloning if the number of output (approximate) clones tends to be very large. Here, we review some selected topics of applications of quantum states discrimination.

\subsection{Pusey-Barrett-Rudolph (PBR) theorem}

Although quantum mechanics formulated with its axioms and postulates serves as a good description of microscopic world, the interpretation of quantum states has surprisingly remained an issue of debate: do quantum states represent merely {\it states of knowledge} only or, are they  {\it real physical states}?  It has been a long debate since the beginning of quantum theory, going perhaps as far back as the celebrated article by Einstein, Podolsky, and Rosen \cite{ref:epr} and also the results of Bell \cite{ref:bell}. Recently, a no-go theorem by Pusey, Barrett, and Rudolph \cite{ref:pbr} says that {\it if quantum state merely represents information about the real physical state of a system, then experimental predictions that contradict those of quantum theory can be constructed} \cite{ref:pbr}. Note that the theorem is valid under the assumption that two quantum systems at different locations can be  prepared independently.

A state discrimination method called quantum state exclusion is useful for undertanding  the proof of the theorem. Note that it does not correspond to minimum-error nor unambiguous state discrimination. To derive the theorem, let us assume a physical theory where a system is described by a physical state $\lambda$ or, more generally, its distribution $\mu(\lambda)$. For instance, for classical theory, it is a distribution over the phase space. If a quantum state represents a physical state, corresponding distributions $\mu_{\psi_0} (\lambda)$ and $\mu_{\psi_1} (\lambda)$ of different states $|\psi_0\rangle$ and $|\psi_{1}\rangle$ must be disjoint. The hypothesis is that a quantum state is regarded as merely a state of knowledge, in which distributions of $\mu_{\psi_0} (\lambda)$ and $\mu_{\psi_1} (\lambda)$ contain an overlapping area in the preparation stage. Let $q$ denote the probability that the preparation is performed on the support. After measurement is performed, Bob then makes an error in distinguishing two states with a probability not less than $q$. 

The argument works with two states $|\psi_0\rangle = |0\rangle$ and $| \psi_1 \rangle= |+\rangle = (|0\rangle + |1\rangle) / \sqrt{2}$. Let us assume two systems prepared independently, $\{ |0\rangle |0\rangle, |0\rangle |+\rangle , |+\rangle |0\rangle ,|+\rangle |+\rangle  \}$. With states $\lambda_1$ and $\lambda_2$ for the two systems, the preparation can be described as $\mu_{\psi_0} (\lambda_1) \mu_{\psi_0} (\lambda_2)$, $\mu_{\psi_0} (\lambda_1) \mu_{\psi_1} (\lambda_2)$, $\mu_{\psi_1} (\lambda_1) \mu_{\psi_0} (\lambda_2)$, $\mu_{\psi_1} (\lambda_1) \mu_{\psi_1} (\lambda_2)$. If both systems are prepared in the overlapping support, then the state is compatible with the four quantum states. In other words, measurement does not give null outcome for any of the four states. As it is shown \cite{ref:pbr}, quantum theory allows for the existence of measurement in which one of the four outcomes never shows a detection event. This leads to a contradiction to what has originally been assumed about the interpretation of quantum states. To generalise the argument to arbitrary two states, $n$ quantum systems have been taken and measurement is applied to their $2^n$ states. In the proof of the no-go theorem, the main task is to seek the measurement that give null outcome to one of input states, and invoke a quantum state exclusion principle.


\subsubsection*{Quantum state exclusion}

The method of {\it quantum state exclusion}, first introduced in Ref. \cite{ref:caves}, aims to reject some of the states and reduce the original state space to a subset. Recently, the problem has been formalised in the form of semidefinite programming \cite{ref:exclusion}. For state preparation $\{ q_i ,\rho_i  \}_{i=1}^N$, in contrast to minimum-error state discrimination, one seeks POVM elements $\{ M_{i}\}_{i=1}^N$ such that $\tr[\rho_i M_i]$ is minimised on average over all $i = 1, \cdots, N$. Ideally, it is aimed that $\tr[ \rho_i M_i] =0$ for all $i=1,\cdots, N$. It is worth mentioning how this can be seen in experiments: if one of states, say $\rho_i$, is repeatedly generated and sent for measurement, its corresponding output port $M_i$ never shows detection events even though the other ports may show detection events with probabilities $\tr[M_j \rho_i]$ for $j\neq i$. Repetition of the experiment gives rise to an output port with no detection events successfully excluding a particular state.  

Quantum state exclusion can also be expressed in the form of semidefinite programming \cite{ref:exclusion}. The primal problem can be written as 
\bea
\mathrm{(Primal)} && \min ~~ \sum_{i} q_i \tr[M_i \rho_i] \nonumber \\
&& \mathrm{subject~to ~}~~ \sum_{i} M_i = I,~~ M_i \geq 0,~~ i=1,\cdots, N, \nonumber
\eea
and its dual problem is also easily derived as
\bea
\mathrm{(Dual)} && \max ~~  \tr [ K ]   \nonumber \\
&& \mathrm{subject ~ to ~} ~~ K \leq q_i \rho_i,~~ i = 1,\cdots, N. \nonumber 
\eea
The above program can be compared to minimum-error state discrimination in Eqs. (\ref{eq:primal}) and (\ref{eq:dual}).  Technically, the two primal problems achieve optimisation in the opposite directions. 
To derive the PBR theorem, two distinct states $|\psi_{k} \rangle = \cos (\theta/2) |0\rangle + (-1)^{k} \sin (\theta/2) |1\rangle$ with $k=0,1$ are considered, and  quantum state exclusion is applied to the product states 
\bea
|\Psi (x_1, x_2,\cdots , x_n)  \rangle =  |\psi_{x_1} \rangle \otimes |\psi_{x_2}\rangle  \otimes \cdots \otimes |\psi_{x_n}\rangle 
\eea
where $x_i = \{ 0,1 \}$ for $i=1,\cdots,n$.  By applying entangled POVM, measurement allowing for state exclusion is optimised over $\H^{\otimes n}$.

\begin{figure}[t]
\begin{center}
\includegraphics[width=5in]{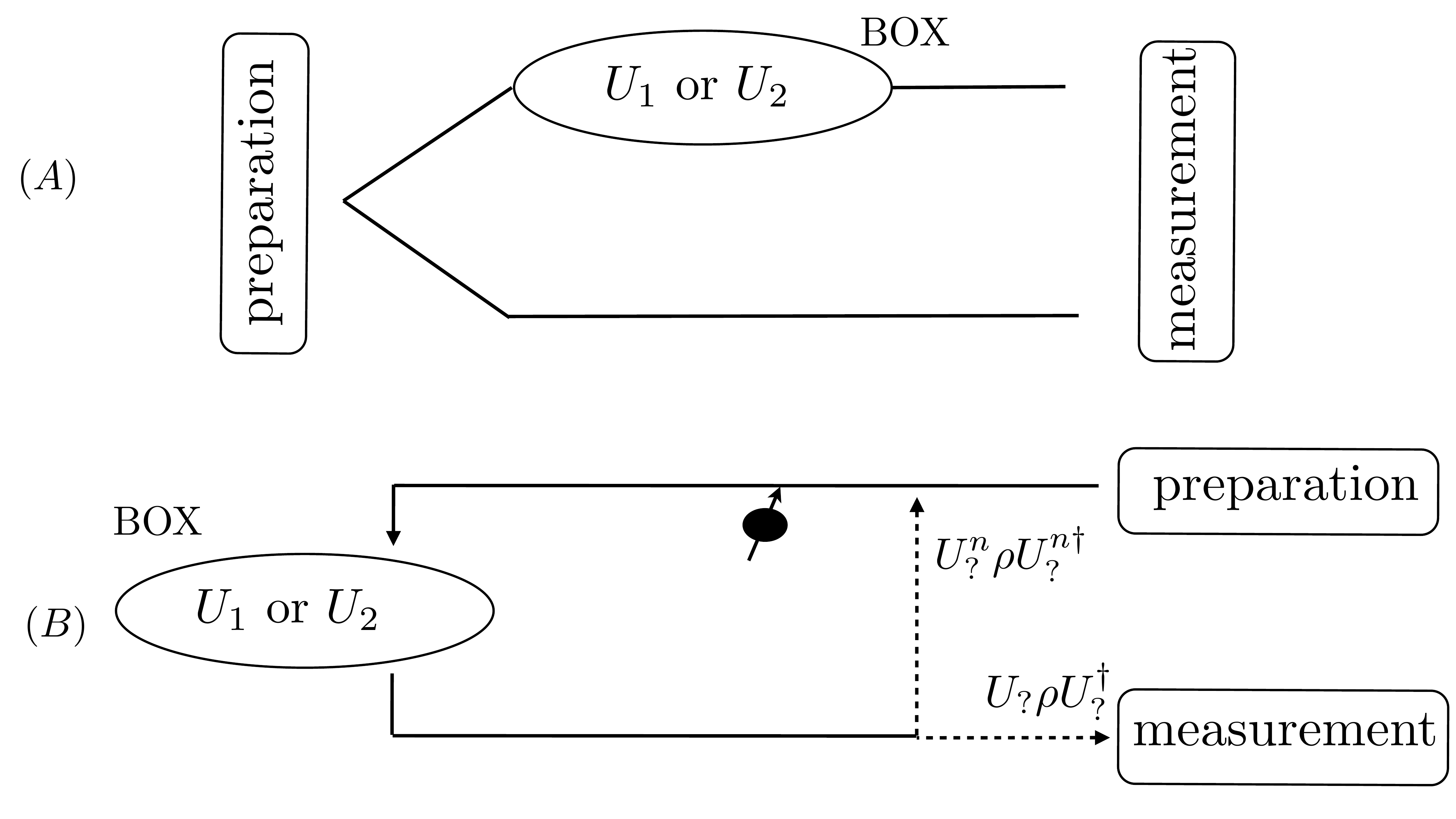}
\caption{ Two schemes of discriminating between two unitaries $U_1$ and $U_2$ are shown. In (A), ancillary systems are introduced and it corresponds to discrimination between $U_1 \otimes I$ and $U_2 \otimes I$. Differently to the case of state discrimination, ancillary systems enhance the distinguishability. In (B), it is allowed to exploit the box many times. After $n$ repetitions, it corresponds to discrimination between $U_{1}^{n}$ and $U_{2}^{n}$. Contrast to state discrimination, for any two unitaries there exist a finite $n$ and input state $\rho$ such that two unitaries after $n$ repetitions are perfectly distinguishable. }
\label{fig5:fig5}
\end{center}
\end{figure}

\subsection{Discrimination of unitary transformations}

Suppose that there is a box in which a unitary transformation, say one of the two unitaries $U_1$ or $U_2$, is applied with some probabilities, say $1/2$, respectively. How then can one learn about which transformation is applied?  A natural approach is to prepare and send a fixed quantum state $|\psi\rangle$ into the box and then discrimnate between $U_1 |\psi\rangle$ or $U_2 |\psi \rangle$, corresponding to two-state discrimination. Unlike the usual two-state discrimination, there is a freedom to choose the input state so that the two ``output" states are better distinguishable. One may then argue that discrimination between unitaries is therefore equivalent to the case of quantum states. The argument might come from the close connections, or similarity, between states and dynamics: for instance, for two distinct quantum states, there always exist a fixed state and two unitaries by which the two states are prepared. However, it turns out that this intuition does not work in a discrimination task. 

From Eq. ({\ref{eq:helbd}}), the distinguishability is quantified by the trace distance between two resulting states, and thus the optimisation is given by
\bea
u = \max_{\rho} \|  U_1 \rho U_{1}^{\dagger} -   U_2 \rho U_{2}^{\dagger} \|_1 \label{eq:unid1}
\eea
with the guessing probability, $p_{\g} = (1 + u/2) /2$. \\

Recall that the main goal here is not state discrimination but discrimnating one of the two unitary transformations.  To this end, one can exploit ancillary systems, that is, discrimination between $I\otimes U_1$ and $I \otimes U_2$ \cite{ref:dariano}, see Fig. (\ref{fig5:fig5}). Assuming that the ancillary system has the dimension as the system, the optimisation task becomes, for an input state $\rho\in S(\H \otimes \H)$,
\bea
u  = \max_{\rho} \| I\otimes U_1 \rho I\otimes U_{1}^{\dagger} - I\otimes  U_2 \rho I\otimes U_{2}^{\dagger} \|_1. \label{eq:unid2}
\eea
As an example, we consider two unitary transformations, say Pauli operators $X$ and $Z$. Note that for any input state $|\varphi\rangle$, it holds that $\langle  \varphi |  X Z  |\varphi \rangle \neq 0$, showing that they are not perfectly distinguishable without ancillary systems, see Eq. (\ref{eq:unid1}).  Let us now consider an input state $|\psi\rangle = (|00\rangle + |11\rangle )/\sqrt{2}$.  It is easily shown that 
\bea
\langle \psi | (X \otimes I) (Z \otimes I)  | \psi\rangle = 0. \nonumber 
\eea
This shows that we have $u=2$ in Eq. (\ref{eq:unid2}) and $p_{\g}=1$. That is, the two non-commuting unitary transformations are perfectly distinguishable by introducing ancillary systems. Note that the two corresponding states are not perfectly distinguishable even though ancillary systems are considered. \\


{\it Discrimination in the asymptotic limit.}  Suppose now that one can repeatedly exploit the box involving the unitary transformations, i.e.  with the input state as $|\psi\rangle$, the box applies one of two unitary transformation $U_1$ or $U_2$ repeatedly many times so that, after $n$ repetitions, resulting states are either $U_{1}^n |\psi\rangle $ or $U_{2}^n |\psi\rangle $.  To guess which unitary transformation is used, one looks at the two-state discrimination of resulting states. The problem is precisely to find an input state $|\psi\rangle$ such that the resulting two states $U_{1}^n |\psi\rangle $ and $U_{2}^n |\psi\rangle $ are orthogonal to each other. The task is therefore
\bea
\max_{\rho} \|  U_1^{n} \rho U_{1}^{n \dagger}  - U_2^{n} \rho U_{2}^{n \dagger} \|. \nonumber 
\eea
In Ref. \cite{ref:acinu}, it is shown that there always exist finite $n$ and an input state $\rho$ such that two resulting states $U_1^{n} \rho U_{1}^{n \dagger}$ and $U_2^{n} \rho U_{2}^{n \dagger}$ become orthogonal, that is, perfectly distinguishable. In the case of state discrimination, two non-orthogonal  states are never perfectly distinguishable with certainty even if a number of copies are provided. Hence, there is intrinsic difference between state discrimination and the discrimination of unitary transformations.

\begin{figure}[t]
\begin{center}
\includegraphics[width=5in]{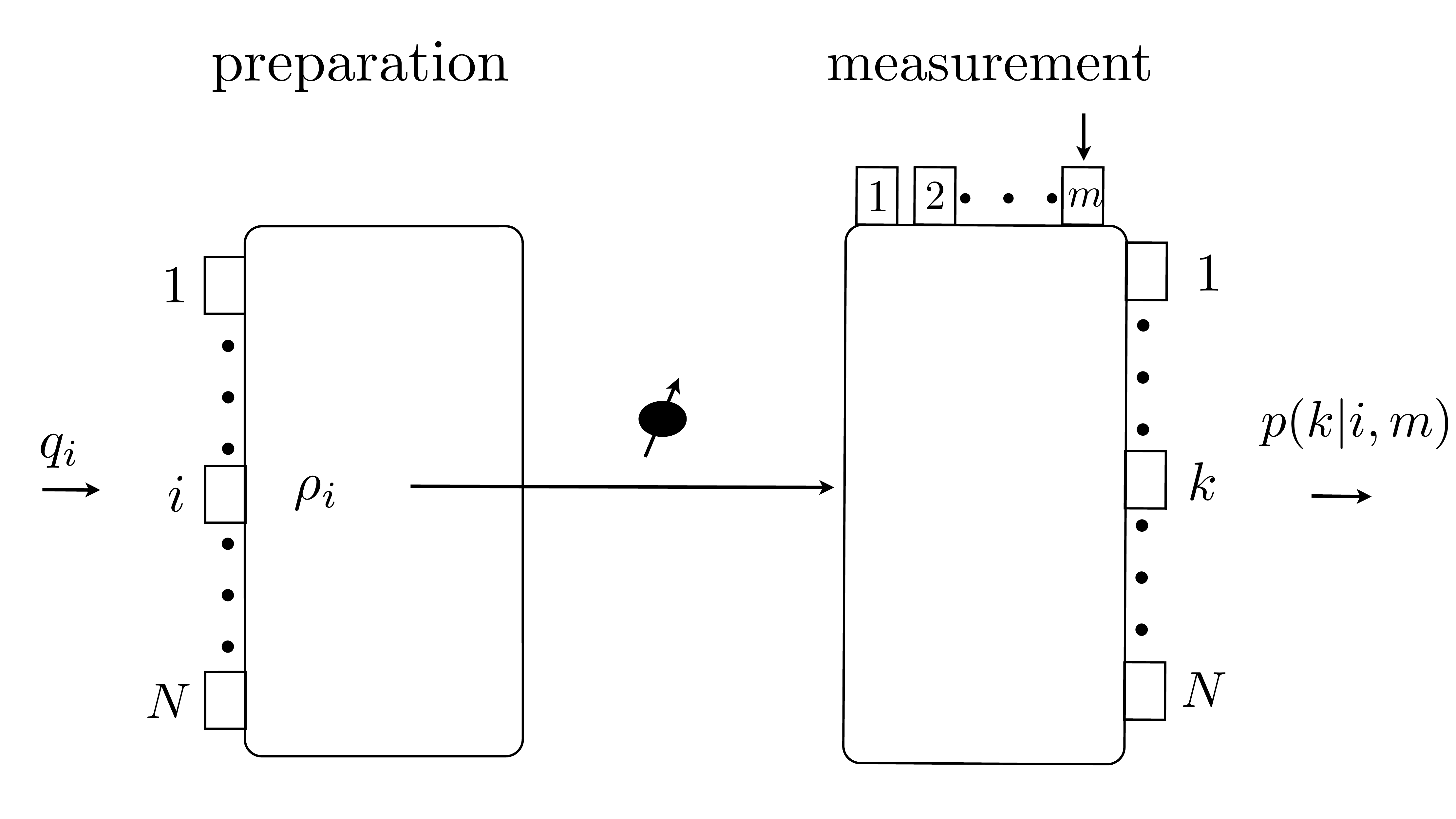}
\caption{A setting for dimension witness using quantum state discrimination is shown. In this case, no {\it a priori} information is known about state preparation: pressing one of $N$ buttons, say $i$, corresponds to generation of unkonwn state $\rho_i$, and measurement is applied by a setting $m$ giving an outcome $k$. The probability that this happens is denoted by $p(k | i,m)$, with which the minimal dimension of Hilbert space describing states $\{\rho_i \}_{i=1}^N$ can be then certified. }
\label{fig6:fig6}
\end{center}
\end{figure}

\subsection{ Dimension witnesses }


For dimension witnesses, one looks at the question: in an experiment without  full knowledge of the measurement and preparation of unknown states, what features of the system can be gleaned from observed data? In Fig. (\ref{fig6:fig6}), the experimental setup is depicted where the observed data refers to the collection of probabilities $p(k|i,m)$ of having outcome $k$ when a state $\rho_i$ is prepared by pressing button $i$ with measurement $m$. It turns out that minimal dimensions of classical and quantum systems can be certified by observed data. A general method has been presented in Ref. \cite{ref:dimwit}, based on numerical tests. 


Recently, it is shown that quantum state discrimination can also be used to certify the minimum dimension of a Hilbert space of a given quantum system \cite{ref:dimqsd}. Since two-state discrimination is by far the only case where optimal discrimination is completely analyzed for all dimensions, it can be applied in the context of dimension witnesses. We first specify the POVM elements so that $M_{k}^{m}$ to denote the operator giving outcome $k$ when measurement $m$ is chosen. For complete measurement, it holds that $\sum_{k} M_{k}^m = I$. In Ref. \cite{ref:dimqsd}, the are $N$ possible preparations and $M$ possible measurements with $M = N (N-1)/2$ so that each measurement consists of two outcomes $ k \in \{- 1,1\}$ only. Indeed, each measurement can be labeled by $m = (x,x')$ for $x,x' \in \{1,\cdots, N  \} $. The witness is then bounded by
\bea
W_N = \sum_{x > x'} |  p(k=1 | x, (x,x') ) - p(k=1 | x' , (x,x'))  |^2 \leq Q_d = \frac{N^2}{2} (1 - \frac{1}{\min(d,N)}).  \nonumber
\eea
Let us explain the above equation. First, the difference between two probabilities corresponds to the trace distance between two unknown states $\rho_x$ and $\rho_{x'}$ among $x,x' \in \{1,\cdots, N \}$, that is, the minimum-error discrimination between the two states. The witness can be understood as the average fluctuations of pair-wise distinguishability of the preparation process. The inequality is fulfilled if the unknown states are described by a Hilbert space of dimension equal to or less than $d$. The inequality would be violated if the quantum dimension in the preparation is greater than $d$. For instance, the upper bound for $W_7$ can be analytically computed in Ref. \cite{ref:dimqsd}: \\
\begin{center}
\begin{tabular}{ccccccccccc}
\hline
 dimension d & \vline & 2&  3 &  4 & 5 &  6 & 7  \\
\hline
 $Q_d$ & \vline & 12.25 & 16.33 & 18.38 & 19.60 & 20.42 & 21 \\
\hline
\end{tabular}\\
\end{center}
The usefulness of applying state discrimination to dimension witnesses, compared to the general formalism in Ref. \cite{ref:dimwit}, lies with the fact the the quantity $Q_d$ can be analytically computed. Despite the advantage, it remains an open question how one could generalize the method to multiple state discrimination and dimension witnesses due to the lack of solutions in general situation.

\subsection{ Equivalence to asymptotic quantum cloning}
\label{subsec:cloning}

Quantum cloning allows the distribution of unknown quantum states from one party to many parties \cite{ref:bh}. Formally, it corresponds to the dynamics from given states $|\phi\rangle $ with additional free resources $|0\rangle^{\otimes M-1}$ to a $M$-partite state $\rho_{\phi}^{(\times M)} \in \B(\H^{\otimes M} )$ so that the resulting state is as close to as $M$ copies of initial state. One may be interested in maximising the fidelity of individual clones. It is called {\it local fidelity} in the context of symmetric cloning where all clones are identical, i.e. for a resulting $M$-partite state $\rho_{\phi}^{(\times M)} \in \B(\H^{\otimes M} )$, it holds that an individual clone $\rho_i = \tr_{\bar{ i }} \rho_{\phi}^{ ( \times M ) } \in \B(\H)$ where $\tr_{\bar{ i }} = \tr_{1, \cdots, i-1, i+1, \cdots, M}$ is identical the others: $\rho_i= \rho_j$ for all $j \neq i$. Thus, the local fidelity computes the fidelity between the input state and an approximate output clone,
\bea
F_L = F (|\phi\rangle , \rho_j) = \langle \phi | \rho_j | \phi \rangle\nonumber
\eea 
where $F$ denotes the fidelity. 

It is interesting to see that state discrimination can be regarded as a form of cloning: for states $\{q_i ,\rho_i \}_{i=1}^N$, the minimum-error discrimination find out which state has been prepared and then produced  as many as copies of the state as one wishes.  In some sense, it is like a measurement-and-preparation scheme in which the information of the unknown quantum state is converted into classical information through measurements and then the quantum state is prepared using the classical information. Of course, the discrimination task is never perfect, and thus the error contributes to  imperfect clones. Note that the situation is exactly a $1\rightarrow \infty$ symmetric cloning, or also called asymptotic cloning. 

A fair comparison of the two processes of cloning, one of optimal quantum cloning and the other based on state discrimination, is made in cases of asymptotic cloning. Indeed, for some cases it is shown that the two processes are equivalent \cite{ref:uniclone, ref:keyl, ref:phaseclone}, and the case of equivalence in general is  an open problem \footnote{This is 22nd problem in the list, http://qig.itp.uni-hannover.de/qiproblems.}. In Refs. \cite{ref:baeacinclone, ref:chiribella}, it is shown that quantum state discrimination, or estimation in the case that given states contain continuous spectrum, is equivalent to optimal asymptotic quantum cloning.


\section{  Characterizations in quantum communication}  
\label{sec:characterization}

One of the main motivations that quantum state discrimination was investigated in the very beginning was due to the huge development in optical communication. As the size of systems in which information is encoded gets smaller, it is natural to consider quantum systems in the limit of the physical size. In quantum communication, the framework relies on quantum postulates: measurement results are interpreted according to the laws of quantum mechanics. Quantum sources also contain entanglement that shows correlations stronger than classical ones. All these are consistent to other physical principles of classical theories. For instance, entanglement does not violate the relativistic causality, or also called as the no-signaling principle that rules out instantaneous communication.  

Interestingly, the formalism of quantum state discrimination is much related to constraints given by other physical principles. Here, we consider two contexts of quantum communication that have tight relations to, and thus also characterize, optimal quantum state discrimination. Both of them consider a cryptographic scenario.


\subsection{ Quantum min-entropy}
\label{sec:min-entropy}


In quantum cryptographic protocols, min-entropy is exploited to quantify privacy amplification that decouples legitimate parties from an eavesdropper. It estimates the number of bits to be sacrificed during privacy amplification protocols \cite{ref:rennerthesis}. It is also a measure that quantifies the randomness of sources. In Ref. \cite{ref:ckr}, an operational meaning of quantum min-entropy has been obtained that it is indeed equivalent to the guessing probability of the minimum-error state discrimination. The precise relation is provided as follows. 

Let us write an entangled state shared by Alice and Bob as $|\psi\rangle_{AB}$. Suppose that Alice applies a measurement with POVM elements $\{  M_{i} \}_{i=1}^N$. On measurement, she gets the outcome, say $j$ from a detection event on POVM $M_j$, and then Bob is left with corresponding quantum state,
\bea
\rho_{B}^{(j)} = \frac{ 1 }{  q_j }    \tr_{A}[ ( (M_j)_A \otimes I_B) |\psi\rangle_{AB} \langle \psi | ]   \label{eq:bobstate}
\eea
where $q_j = \tr[ ( (M_j)_A \otimes I_B) |\psi\rangle_{AB} \langle \psi |  ]$ denotes the probability that Alice has outcome $j$. Bob's states are normalized. Once Alice announces to Bob that she has performed her measurement, Bob knows that he has quantum states $\{\rho_{B}^{(j)} \}_{j=1}^N$.  The state shared between them can now be described by classical-quantum correlations,
\bea
\rho_{AB} = \sum_{j=1}^N q_{j} | j\rangle_A \langle j| \otimes \rho_{B}^{(j)} \label{eq:cqstate}
\eea
where Alice's measurement outcome of $M_j$ is denoted by $| j \rangle_A \langle j|$ for $j=1,\cdots,N$. Note that this state can also be prepared as in subsection\ref{subsec:setting} without shared entanglement \cite{ref:bbm}. 

The classical-quantum state in Eq. (\ref{eq:cqstate}) precisely describes the physical situation  before Bob's measurement.  By applying optimal measurement,  Bob can gain information from his quantum states with the conditional min-entropy
\bea
H_{\min} (X | B) =  - \inf_{\sigma_B} \inf_{\lambda} \{  \lambda \|  \rho_{AB} \leq 2^{\lambda} (I_A \otimes \sigma_B) \} \nonumber
\eea
where the infimum is taken over all states $\sigma_B \in \B (\H)$. Note that the above corresponds to a convex optimisation task.

It turns out that the conditional min-entropy in the above is equivalent to the minimum-error state discrimination \cite{ref:ckr}. More precisely, the equivalence is established with the guessing probability of the minimum-error state discrimination for $\{q_j,\rho_{B}^{(j)} \}_{j=1}^N$ with $\rho_{B}^{(j)}$ in Eq. (\ref{eq:bobstate}) with
\bea
H_{\min} (X|B) = -\log p_{\g}  \label{eq:min-entropy}
\eea
where the explicit form of the guessing probability $p_{\g}$ can be found in Eq. (\ref{eq:medprob}) in the subsection \ref{subsec:med}. The equivalence means that, for Bob to maximise his information, the optimal measurement he chooses should be identical to the optimal one for the minimum-error state discrimination. It also means that optimal state discrimination is a corresponding operational task when min-entropy is applied in cryptographic applications.

\subsection{ Optimal discrimination constrained by the no-signaling principle}

Let us reconsider the bipartite scenario above where the two parties share an entangled state $|\psi\rangle_{AB}$, Alice applies measurement, and Bob is then left some quantum states. Here, we consider a generalization that Alice can choose one of $N$ possible measurement denoted by $\{ M_i^{(A)}\}_{i=1}^N$, where each measurement by Alice $M_i^{(A)}=\{ M_i , I - M_i \} $ is complete and has two outcomes of $M_i$ and $I - M_i$. Since each measurement is complete, as long as Alice does not announce her measurement, Bob does not learn Alice's choice from his quantum states. This already excludes the possibility of instantaneous communication \cite{ref:herbert}.  

Suppose that measurement $M_{i}^{(A)}$ for some $i=1,\cdots, N$ is appled and Alice does not  announce her measurement outcomes. Bob is then left with a mixture of two possibilities, either $\rho_i$ or $\sigma_i$ so that
\bea
\rho_{i}^{(B)} = p_i \rho_i + (1-p_i) \sigma_i,~~\mathrm{for}~i=1,\cdots, N, \label{eq:ens}
\eea
where we have introduced $\rho_{i}$ as the resulting state of Alice's measurement $M_{i}$ and $\sigma_i$ as the state of measurement $I-M_i$,
\bea
\rho_i & =& \frac{1}{p_i} \tr_A [|\psi\rangle_{AB}\langle \psi| M_i \otimes I  ], 
 \label{eq:stateb1}\\
\sigma_i & =& \frac{1}{1-p_i} \tr_A [|\psi\rangle_{AB}\langle \psi| (I -M_i) \otimes I  ],~~\mathrm{where}~ p_i =  \tr [|\psi\rangle_{AB}\langle \psi| M_i \otimes I  ].  \label{eq:stateb2}
\eea
Note that, since Alice's measurement $M_{i}^{(A)}$ is complete for all $i=1,\cdots, N$, it holds that 
\bea
\rho_i^{(B)} = \rho_j^{(B)} ~~\mathrm{for~all} ~i,j=1,\cdots, N. \nonumber
\eea
Bob's strategy is based on the fact that each state $\rho_j$ is contained in the ensemble of $\rho_{j}^{(B)}$ in Eq. (\ref{eq:ens}) as a result of Alice's measurement $M_{j}^{(A)}$. He then prepares his measurement to discriminate the states $\{ \rho_{i} \}_{i=1}^N $, and concludes Alice's measurement from the outcome: he would conclude Alice's measurement $M_{j}^{(A)}$ if a detection event on his side happens in the $j$-th port that finds $\rho_j$ among states $\{\rho_j \}_{j=1}^N. )$. It is clear that, by the no-signaling principle, the strategy would not add any information to Bob: this means that the distinguishability for states $\{ \rho_{i} \}_{i=1}^N$ cannot be arbitrarily precise. 

Bob's discrimination task is restricted to the states $\{\rho_{i} \}_{i=1}^N$ in the ensemble; other states $\{\sigma_i \}_{i=1}^N$ in the ensemble in Eq. (\ref{eq:ens}) may still contribute to detection events, and are considered as errors in the Bob's guessing strategy. In any case, since Bob aims to guess Alice's measurement by discriminating states $\{\rho_{i} \}_{i=1}^N$ only, detection events corresponding to the states $\{\sigma_i \}$ are disregarded in Bob optimal measurement setting. Therefore, the {\it a priori} probabilities exploited for Bob's measurement are 
\bea
q_ i = \frac{\mathrm{Prob} [ \rho_i ] }{\mathrm{Prob} [\{\rho_{i} \}_{i=1}^N ] } = \frac{p _i }{p_1 + \cdots + p_N} \nonumber
\eea
with $\{ p_i \}_{i=1}^{N}$ in Eq. (\ref{eq:stateb2}), where $\mathrm{Prob}[ \rho] $ denotes the probability that $\rho$ appears. Bob's measurement is optimised for the state preparation $\{ q_i ,\rho_{i} \}_{i=1}^N$.  If Bob's guessing probability is so high and he correctly guesses Alice's measurement better above random guesses, a faster-than-light communication would be established.  thus, the guessing probability is constrained by the no-signaling principle. It can be shown that the guessing probability is bounded by
\bea
p_{\g} \leq p_{\g}^{(NS)} = \frac{1}{p_1 + \cdots + p_N}. \label{eq:bdns}
\eea
In other words, violation of the above inequality would lead to instantaneous communication. In fact, the inequality turns out to be tight. That is, for any set $\{ q_i ,\rho_{i} \}_{i=1}^N$, the optimal state discrimination always attains the equality in Eq. (\ref{eq:bdns}) as dictated by the no-signaling principle.

Since analytical expression for optimal state discrimination remains an open problem to date, it is not possible to compare it directly with the upper bound for the strategy. Instead, one can find from the LCP formalism of optimal state discrimination in Eq. (\ref{eq:lcp}) that for any set $\{ q_i ,\rho_i \}_{i=1}^{N}$, there exists a symmetry operator $K$ that is decomposed into $K = q_i \rho_i + r_i \sigma_i $ for all $i=1,\cdots,N$. Normalising the symmetry operator gives 
\bea
\widetilde{K} = \frac{K}{ \tr[K]} = p_i \rho_i + (1-p_i) \sigma_i \nonumber
\eea
and this operator can be related to state $\rho_{i}^{(B)}$ by identifying $p_i = q_i / \tr[K]$ for all $i=1,\cdots,N$ in Eq. (\ref{eq:ens}).  Recall that the guessing probability is given by $\tr[K]$ in Eq. (\ref{eq:trK}). With the fact that $\sum_{i}q_i =1$, we have the guessing probability 
\bea
p_{\g} = \frac{1}{p_1 + \cdots + p_n} \nonumber
\eea
which is equal to the upper bound in Eq. (\ref{eq:bdns}). It is also worth emphasising that the scenario constraining the no-signaling principle can be exploited to discover optimality conditions in Eq. (\ref{eq:lcp}). 

The scenario shown  above is generic in many quantum communication protocols.  Imposing the impossibility of a faster-than-light communication is a useful tool to probe the capabilities of manipulating quantum states, for instance, quantum cloning \cite{ref:gisin98} or local quantum dynamics \cite{ref:simon}. In fact, optimal quantum cloning is also tightly related to the no-signaling principle \cite{ref:gisin98}: a faster-than-light communication would happen if cloning of quantum states could work better than optimal quantum cloning. As mentioned in the subsection \ref{subsec:cloning}, quantum state discrimination is a form of quantum cloning, and therefore one may expect relations between the no-signaling principle and state discrimination. Indeed, optimal quantum state discrimination is also be tightly characterized by the no-signaling principle.

\section{Conclusion}
\label{sec:conclusion}

Quantum state discrimination serves as a basic tool for both quantum information theory and the foundation of quantum mechanics. Although general theorems regarding optimal state discrimination remains unsolved, much progress has been gained in recent years on some special cases. The technical difficulty in most general scenario may place strict limitations on the development of some quantum information tasks. Moreover, since the quantum state discrimination is closely related to some existing hard problems, developments in this direction could lead to new perspectives and challenges. The present review has provided a comprehensive introduction to quantum state discrimination and its selected applications. Although it is not covered here, there has also been significant progress in discrimination of continuous variable states, see for instance, Refs. \cite{ref:becerra1, ref:becerra2}, for recent experimental implementation along the line. They are in fact related to applications such as quantum key distribution with continuous variable systems, e.g. \cite{ref:qkdcon}. There are also excellent review articles, e.g. Refs. \cite{ref:rev1, ref:rev2, ref:rev3, ref:rev4, ref:rev5}, on quantum state discrimination from different angles. The present article mainly focus on recent development in the methods of state discrimination from the theoretical point of view and its selected applications. It is hoped that these new results in quantum state discrimination may lead to new insights for other issues in quantum information science and vice versa.

\section*{Acknowledgement}

The work was done partially while the authors were visiting the Institute for Mathematical Sciences, National University of Singapore in 2013, being supported by the Institute. This work is supported by the People Programme (Marie Curie Actions) of the European Union's Seventh Framework Programme (FP7/2007-2013) under REA grant agreement N. 609305, and the National Research Foundation \& Ministry of Education, Singapore. 






\end{document}